\documentclass[]{article}
\usepackage{amsmath}
\usepackage{amsfonts}
\usepackage{amssymb}
\usepackage{siunitx}
\usepackage{graphicx}
\usepackage{physics}
\usepackage[affil-it]{authblk}
\usepackage{appendix}

\title{Super Heat Kernel and One-Loop Divergence of Super Yang-Mills Theory in Conformal Supergravity\thanks{Work supported in part by the Director, Office of Science, Office of High Energy and Nuclear Physics, Division of High Energy Physics, of the US Department of Energy under Contract DE-AC02-05CH11231 and in part by the National Science Foundation under grant PHY-1316783.}}
\author{Ka-Hei Leung\thanks{Email: kaheileung@berkeley.edu}}
\affil{Department of Physics and Theoretical Physics Group,\\
	Lawrence Berkeley National Laboratory, University of California, \\
	Berkeley, California 94720, USA}
\date{\vspace{-4ex}}

\begin{document}

\maketitle

\begin{abstract}
We consider super Yang-Mills Theory in $N=1$ conformal supergravity. Using the background field method and the Feddeev-Popov procedure, the quantized action of the theory is presented. Its one-loop effective action is studied using the heat kernel method. We shall develop a non-iterative scheme, generalizing the non-supersymmetric case, to obtain the super heat kernel coefficients. In particular, the first three coefficients, which govern the one-loop divergence, will be calculated. We shall also demonstrate how to schematically derive the higher order coefficients. The method presented here can be readily applied to various quantum theories. We shall, as an application, derive the full one-loop divergence of SYM in conformal supergravity. 
\end{abstract}

\newpage
\section{Introduction}\label{intro}
Supergravity has been a subject of interest over the past thirty years. It is believed that it could represents the low-energy effective model of a more fundamental theory of quantum gravity, for instance string theory. Therefore despite being not renormalizable similar to its non-supersymmetric counterpart, understanding supergravity at the quantum level may provide insights on how qunatum gravity behaves away from Planck scale. In particular, supergravity at the one-loop level was extensively studied as it captures the majority of the features of quantum supergravity. A considerable amount of work has been done on topics like one-loop divergences, regularization, and anomalies, some examples being \cite{Gaillard94}, \cite{Gaillard97}, \cite{Gaillard99} and \cite{Butter15}.

A natural extension of supergravity is to introduce conformal symmetry, and the class of models with this extra symmetry is naturally called conformal supergravity. Historically it was introduced by Kaku, Townsend, van Nieuwenhuizen, Ferrara and Grirasu \cite{Kaku78a,Kaku78b,Townsend79,Ferrara78}. Conformal supergravity was initially developed in the component approach. But Butter \cite{Butter10} later discovered it can be described as a theory with superfields in conformal superspace. It is known that one can reduce the theory to the ordinary Poincare supergravity, or $U(1)_K$ supergravity developed by Bin\'etruy, Girardi and Grimm \cite{BBG00}, by choosing a special gauge. Hence by studying conformal supergravity, which seemed to be a restricted class of theories, one can also understand supergravity in general.

In the following, we shall consider super Yang-Mills theory in conformal supergravity, which by gauge fixing covers the case of usual Poincar\'e supergravity as well. We will examine its one-loop effective action via the heat kernel method, first developed by de Witt \cite{DeWitt64}. The central idea is to have a series expansion of the quantum heat kernel, with certain coefficients of the expansion encoding one-loop divergences of the theory. It is possible to evaluate such coefficients starting from the lowest order by either a recursive method, as was done originally by de Witt, or via non-recursive means, for example methods by Avramidi \cite{Avramidi91}. However, most of the work done has been on non-supersymmetric theories. In this work we shall describe how to extend Avramidi's method to the case of supersymmetric theories. This allows us to calculate super heat kernel coefficients up to any order non-iteratively, more effective than for instance the recursive method done in \cite{Buchbinder86}, or the Fourier transform method in \cite{McArthur83,McArthur84}.

This work is divided into two parts. In the first we shall discuss the quantization of the super Yang-Mills theory in conformal supergravity. The quantized quantum action, which will be crucial in one-loop studies, will be derived. The second part will be devoted to studying the one-loop effective action and divergence of the theory. By using the heat kernel method, we shall ultimately calculate the full one-loop divergence of the SYM theory.

\section{Conformal Supergravity}\label{conformalSUGRA}
In the following the superspace formulation of $\mathcal{N}=1$ conformal supergravity discovered by Butter \cite{Butter10} will be reviewed, which can be shown to be equivalent to the original component approach. A detailed study of comparison between the two formalisms can be found in \cite{Kugo16a}\cite{Kugo16b}. We shall closely follow the notation and convention of Butter's original paper.

\subsection{Superconformal Algebra and Conformal Superspace}\label{conformalSUGRA1}
We start by constructing the superspace from the super-conformal algebra. Consider first the case without supersymmetry, the Poincar\'e algebra is 
\begin{equation}\label{Poincarealg}
\begin{aligned}
& \left[M_{ab},M_{cd}\right]=\eta_{bc}M_{ad}+\eta_{ad}M_{bc}-\eta_{ac}M_{bd}-\eta_{bd}M_{ac}\\
& \left[M_{ab},P_c\right]=P_a\eta_{bc}-P_b\eta_{ac} \\
& [P_a,P_b]=0.
\end{aligned}
\end{equation}
Note the definition of the generators are so that they are anti-Hermitian, and the structure constants of the algebra are real. For example, the differential representation for $P_a$ is $P_a=\partial_a$. One can extend the algebra to the conformal one by adding the \emph{dilatation} operator $D$ and the \emph{special conformal transformation} $K_a$, the extra commutation relations being: 
\begin{equation}\label{conformalalg}
\begin{aligned}
& \left[D,P\right]=P, \quad [D,K_a]=-K_a, \quad [D,M_{ab}]=0 \\
& \left[M_{ab},K_c\right]=K_a\eta_{bc}-K_b\eta_{ac} \\
& [K_a,P_b]=2\eta_{ab}D-M_{ab}.
\end{aligned}
\end{equation}

Now let us introduce $\mathcal{N}=1$ supersymmetry, by adding the fermionic operator $Q_\alpha$ and its conjugate $\bar{Q}^{\dot{\alpha}}$. The well-known graded commutation relation is, with $(\sigma_{ab})_\alpha^\beta$ being the Lorentz generator in the spinor representation,
\begin{equation}\label{susyalg}
\begin{aligned}
& \left[Q_\alpha,\bar{Q}_{\dot{\alpha}}\right]=-2i\sigma^a_{\alpha\dot{\alpha}}P_a \\
& [M_{ab},Q_\alpha]=(\sigma_{ab})_\alpha^\beta Q_\beta \\
& \left[P_a,Q_\alpha\right]=0.
\end{aligned}
\end{equation}
Here the grading will be implicit as in \cite{Butter10}, thus $[\,,\,]$ is actually the graded commutator: $[A,B]=AB-(-1)^{AB}BA$, with a handy notation being adopted: the exponent in $(-1)^{A}$ is the fermion number of the generator $A$, which equals 0 if $A$ is bosonic and 1 if fermionic. The conformal extension requires two extra operators, the chiral rotation generator $A$, and the fermionic counterpart of $K_a$: $S_\alpha$ and $\bar{S}^{\dot{\alpha}}$. The commutation relations will be:
\begin{equation}\label{conformalsusyalg}
\begin{aligned}
& [D,Q_\alpha]=\frac{1}{2}Q_\alpha, \quad [D,\bar{Q}^{\dot{\alpha}}]=\frac{1}{2}\bar{Q}^{\dot{\alpha}}, \\
& [D,S_\alpha]=-\frac{1}{2}S_\alpha, \quad [D,\bar{S}^{\dot{\alpha}}]=-\frac{1}{2}\bar{S}^{\dot{\alpha}},\\
& [A,Q_\alpha]=-iQ_\alpha, \quad [A,\bar{Q}^{\dot{\alpha}}]=i\bar{Q}^{\dot{\alpha}} , \\
& [A,S_\alpha]=iS_\alpha, \quad [A,\bar{S}^{\dot{\alpha}}]=-i\bar{S}^{\dot{\alpha}}, \\
& [K_a,Q_\alpha]=i\sigma_{a\alpha\dot{\beta}}\bar{S}^{\dot{\beta}}, \quad [K_a,\bar{Q}^{\dot{\alpha}}]=i\bar{\sigma}_a^{\dot{\alpha}\beta}S_\beta, \\  & [M_{ab},S_\alpha]=(\sigma_{ab})_\alpha^\beta S_\beta, \quad \left[S_\alpha,\bar{S}_{\dot{\alpha}}\right]=2i\sigma^a_{\alpha\dot{\alpha}}K_a, \\
& [S_\alpha,P_a]=i\sigma_{a\alpha\dot{\beta}}\bar{Q}^{\dot{\beta}}, \quad [\bar{S}^{\dot{\alpha}},P_a]=i\bar{\sigma}_a^{\dot{\alpha}\beta}Q_\beta, \\
& [S_\alpha,Q_\beta]=(2D-3iA)\epsilon_{\alpha\beta}-2M_{\alpha\beta}, \\ 
& [S^{\bar{\alpha}},\bar{Q}^{\dot{\beta}}]=(2D+3iA)\bar{\epsilon}^{\dot{\alpha}\dot{\beta}}-2M^{\dot{\alpha}\dot{\beta}}.
\end{aligned}
\end{equation}
Other commutation relations not shown above are just zero. Here $M_{\alpha\beta}=(\sigma^{ba}\epsilon)_{\alpha\beta}M_{ab}$ and $M^{\dot{\alpha}\dot{\beta}}=(\bar{\sigma}^{ba}\epsilon)^{\dot{\alpha}\dot{\beta}}M_{ab}$, roughly speaking they are projections of $M_{ab}$ that transform only undotted spinors and dotted spinors repsctively.

The conformal superspace can be constructed by gauging the above algebra. The space has collection of coordinates $z^M=(x^m, \theta^\mu, \bar{\theta}_{\dot{\mu}})$. For each operator in the algebra $X_{\mathcal{A}}$, associated with it is a gauge connection field $W_M{}^{\mathcal{A}}$:
\begin{equation}\label{guagefielddef}
\begin{aligned}
W_M{}^{\mathcal{A}}X_{\mathcal{A}}&= E_M{}^AP_A+\frac{1}{2}\phi_M{}^{ba}M_{ab}+B_MD+A_MA+f_M{}^AK_A\\
&=E_M{}^AP_A+h_M{}^{\underline{\mathcal{A}}}X_{\underline{\mathcal{A}}},
\end{aligned}
\end{equation}
$E_M{}^A$ is the supervielbein, and $X_{\underline{\mathcal{A}}}$ denotes all generators except $P_A$. $h_M{}^{\underline{\mathcal{A}}}$ will be the connection corresponding to the generator, for instance $B_M$ is the connection field for $D$. An important fact is that a connection encodes the infinitesimal gauge transformation $\delta_g$ of a field:
\begin{equation}\label{defgaugeconnection}
\mathcal{L}_\xi\Phi(x)=\delta_g(\xi^MW_M{}^{\mathcal{A}}X_{\mathcal{A}})\Phi(x),
\end{equation}
$\mathcal{L}_\xi$ is the Lie derivative, or equivalently, an infinitesimal general coordinate transformation with parameter $\xi_M$: $\mathcal{L}_\xi=\delta_{\textrm{GC}}(\xi)$. If $\Phi$ is a scalar without any Einstein indices, we have $\mathcal{L}_\xi\Phi=\xi^M\partial_M\Phi$, thus
\begin{equation}\label{scalarvariation}
\partial_M\Phi=E_M{}^AP_A\Phi+h_M{}^{\underline{\mathcal{A}}}X_{\underline{\mathcal{A}}}\Phi.
\end{equation}
This in particular implies that $E_M{}^AP_A$ acts as the covariant derivative:
\begin{equation}\label{P=covd}
E_M{}^AP_A\Phi=\nabla_M\Phi=(\partial_M-h_M{}^{\underline{\mathcal{A}}}X_{\underline{\mathcal{A}}})\Phi,
\end{equation}
and inverting the vielbein gives 
\begin{equation}\label{P=covd2}
P_A\Phi=E_A{}^M\nabla_M\Phi=\nabla_A\Phi. 
\end{equation}
This certainly makes sense as $P_A$ should represents translations.

To introduce curvature to the space, one deforms the conformal algebra such that $[P_A,P_B]$ develops a non-zero commutator, while retaining other commutation relations:
\begin{equation}\label{defromedP}
\begin{aligned}
\left[P_A,P_B\right]&=-R_{AB}{}^{\mathcal{A}}X_{\mathcal{A}} \\
&=-T_{AB}{}^CP_C-\frac{1}{2}R_{AB}{}^{dc}M_{cd}-H_{AB}D-F_{AB}A-R(K)_{AB}{}^CK_C.
\end{aligned}
\end{equation}
Objects like $H_{AB}$, $F_{AB}$ will be the curvature field of the corresponding generator. These curvature terms can be expressed in terms of the connection fields in \eqref{guagefielddef}. Notice that the connection fields transform under an infinitesimal transformation, with parameter $g=g^{\mathcal{A}}X_{\mathcal{A}}$, as 
\begin{equation}\label{connectiontrans}
\delta_{g}W_M{}^{\mathcal{A}}=\partial_Mg^{\mathcal{A}}+W_M{}^{\mathcal{B}}g^{\mathcal{C}}f_{\mathcal{C}\mathcal{B}}{}^{\mathcal{A}}.
\end{equation}
$f_{\mathcal{C}\mathcal{B}}{}^{\mathcal{A}}$ here are the structure constants of the algebra:
\begin{equation}\label{defstructureconst}
[X_{\mathcal{C}},X_{\mathcal{B}}]=-f_{\mathcal{C}\mathcal{B}}{}^{\mathcal{A}}X_{\mathcal{A}}. 
\end{equation}
Recall equation \eqref{defgaugeconnection} relates gauge transformations to coordinate transformations, requiring the consistency relation $[\delta_{\textrm{GC}}(\xi),\delta_{\textrm{GC}}(\chi)]=-\delta_{\textrm{GC}}([\xi,\chi])$ results in
\begin{equation}\label{dWequation}
\partial_MW_N{}^{\mathcal{A}}-\partial_NW_M{}^{\mathcal{A}}-W_M{}^{\mathcal{B}}W_N{}^{\mathcal{C}}f_{\mathcal{C}\mathcal{B}}{}^{\mathcal{A}}=0.
\end{equation}
In other words, the \emph{gauge curvature} of the connection 1-form $W{}^{\mathcal{A}}$ vanishes. By splitting into the generators into $P_A$ and $X_{\underline{\mathcal{A}}}$, we have the expression for $R_{AB}{}^{\mathcal{A}}$ as 
\begin{equation}\label{Requation}
R_{MN}^{\mathcal{A}}=
\partial_MW_N{}^{\mathcal{A}}-\partial_NW_M{}^{\mathcal{A}}-(E_M{}^Bh_N{}^{\underline{\mathcal{C}}}-E_N{}^Bh_M{}^{\underline{\mathcal{C}}})f_{\underline{\mathcal{C}}B}{}^{\mathcal{A}}-h_M{}^{\underline{\mathcal{B}}}h_N{}^{\underline{\mathcal{C}}}f_{\underline{\mathcal{C}}\underline{\mathcal{B}}}{}^{\mathcal{A}},
\end{equation}
\begin{equation}\label{Requation2}
R_{AB}{}^{\mathcal{A}}=E_B{}^NE_A{}^MR_{MN}^{\mathcal{A}}.
\end{equation}
Alternatively in differential form notation, the expression is
\begin{equation}\label{Requation3}
R^{\mathcal{A}}=
dW^{\mathcal{A}}-E^B\wedge h^{\underline{\mathcal{C}}}f_{\underline{\mathcal{C}}B}{}^{\mathcal{A}}-\frac{1}{2}h^{\underline{\mathcal{B}}}\wedge h^{\underline{\mathcal{C}}}f_{\underline{\mathcal{C}}\underline{\mathcal{B}}}{}^{\mathcal{A}}.
\end{equation}

\subsection{Constraints}\label{conformalSUGRA1.3}
The curvature introduced certainly contain too many degrees of freedom to make a sensible theory. Similar to the case of Poincar\'e supergravity, one imposes constraints to reduce the number of dynamical fields appearing. For example, demanding chiral fields exist will impose $\{\bar{\nabla}_{\dot{\alpha}},\bar{\nabla}_{\dot{\beta}}\}=0$. The following are the constraints imposed:
\begin{enumerate}
	\item Torsion constraints:
	\begin{equation}\label{constraint1}
	\begin{aligned}
	&T_{\alpha\beta}{}^C=T_{\dot{\alpha}\dot{\beta}}{}^C=0\\
	&T_{\alpha\dot{\beta}}{}^c=2i\sigma_{\alpha\dot{\beta}}{}^c\\
	&T_{a\underline{\beta}}{}^C=T_{ab}{}^c=0,
	\end{aligned}
	\end{equation}
	where $\underline{\beta}$ means either $\beta$ or $\dot{\beta}$.
	\item Lorentz curvature constraints:
	\begin{equation}\label{constraint2}
	R_{\underline{\alpha}\underline{\beta}}{}^{ab}=0.
	\end{equation}
	\item Chiral curvature and dilatation curvature constraints:
	\begin{equation}\label{constraint3}
	\begin{aligned}
	&F_{\underline{\alpha}\underline{\beta}}=F_{\underline{\alpha}b}=0\\
	&H_{\underline{\alpha}\underline{\beta}}=H_{\underline{\alpha}b}=0.
	\end{aligned}
	\end{equation}
	\item Special conformal curvature constraints:
	\begin{equation}\label{constraint4}
	R(K)_{\underline{\alpha}\underline{\beta}}{}^C=0.
	\end{equation}
\end{enumerate}
Note that the constraints above shows that the covariant derivatives satisfies
\begin{equation}\label{covdericonstraint}
\begin{aligned}
&\{\nabla_{\alpha},\nabla_{\beta}\}=\{\bar{\nabla}_{\dot{\alpha}},\bar{\nabla}_{\dot{\beta}}\}=0\\
&\{\nabla_{\alpha},\bar{\nabla}_{\dot{\beta}}\}=-2i\nabla_{\alpha\dot{\beta}}.
\end{aligned}
\end{equation}
This is analogous to the Yang-mills case of flat superspace.

One has to verify that the above sets of constraints are valid by solving the Bianchi identities. In this case one finds that the curvature can be expressed by a gaugino-like field $\mathcal{W}_{\underline{\alpha}}=\mathcal{W}_{\underline{\alpha}}{}^{\mathcal{A}}X_{\mathcal{A}}$:
\begin{equation}\label{guaginodef}
\begin{aligned}
&R_{\alpha,\beta\dot{\beta}}^{\mathcal{A}}=2i\epsilon_{\alpha\beta}\mathcal{W}_{\dot{\beta}}^{\mathcal{A}}\\
&R_{\dot{\alpha},\beta\dot{\beta}}^{\mathcal{A}}=2i\epsilon_{\dot{\alpha}\dot{\beta}}\mathcal{W}_{\beta}^{\mathcal{A}}\\
&R_{\alpha\dot{\alpha},\beta\dot{\beta}}^{\mathcal{A}}=-\epsilon_{\dot{\alpha}\dot{\beta}}\{\nabla_\alpha,\mathcal{W}_{\beta}{}^{\mathcal{A}}\}-\epsilon_{\alpha\beta}\{\bar{\nabla}_{\dot{\alpha}},\mathcal{W}_{\dot{\beta}}{}^{\mathcal{A}}\}.
\end{aligned}
\end{equation}
Here $R_{\alpha,\beta\dot{\beta}}^{\mathcal{A}}=\sigma_{\beta\dot{\beta}}{}^bR_{\alpha b}$, $R_{\dot{\alpha},\beta\dot{\beta}}^{\mathcal{A}}$ and $R_{\alpha\dot{\alpha},\beta\dot{\beta}}^{\mathcal{A}}$ are similarly defined. \footnote{In fact, the last equation of \eqref{guaginodef} still holds when the generator index $\mathcal{A}$ is absent due to the specific structure of the torsion tensor \cite{Butter10}.} These fields satisfy the following conditions:
\begin{equation}\label{guaginocon}
\begin{aligned}
&\{\nabla_\alpha,\mathcal{W}_{\dot{\beta}}\}=\{\bar{\nabla}_{\dot{\alpha}},\mathcal{W}_{\beta}\}=0\\
&\{\nabla^\alpha,\mathcal{W}_{\alpha}\}=\{\bar{\nabla}_{\dot{\beta}},\mathcal{W}^{\dot{\beta}}\}.
\end{aligned}
\end{equation}
It turns out also that $\mathcal{W}_{\alpha}$ has no $P_A$, $D$ and $A$ components:
\begin{equation}\label{Wcomponent}
\mathcal{W}_{\alpha}=\frac{1}{2}\mathcal{W}(M)_{\alpha}{}^{cb}M_{bc}+\mathcal{W}(K)_{\alpha}{}^{A}K_A,
\end{equation}
and it can be expressed in terms of one single symmetric chiral field $W_{\alpha\beta\gamma}$, the details of which are omitted here. The field $W_{\alpha\beta\gamma}$ satisfies
\begin{equation}\label{Wabcequation}
\begin{aligned}
&\bar{\nabla}_{\dot\alpha}W_{\alpha\beta\gamma}, \quad DW_{\alpha\beta\gamma}=\frac{3}{2}W_{\alpha\beta\gamma}, \quad AW_{\alpha\beta\gamma}=iW_{\alpha\beta\gamma}, \quad K_AW_{\alpha\beta\gamma}=0\\
&\nabla^\beta\nabla^\gamma{}_{\dot{\alpha}}W_{\gamma\beta\alpha}=-\nabla^{\dot{\beta}}\nabla^{\dot{\gamma}}{}_\alpha W_{\dot{\gamma}\dot{\beta}\dot{\alpha}}.
\end{aligned}
\end{equation}
All curvature terms can be expressed in terms of $W_{\alpha\beta\gamma}$ and its conjugate, and the details can be found in the original reference \cite{Butter10}. Thus $W_{\alpha\beta\gamma}$ is the only dynamical field encoding the geometry of the conformal superspace.

\subsection{Conformal Supergravity Action}\label{conformalSUGRA2}
Matter in conformal superspace is described by \emph{primary} superfields. These fields satisfy the following condition:
\begin{equation}\label{prifielddef}
M_{ab}\Phi=S_{ab}\Phi, \quad D\Phi=\Delta\Phi, \quad A\Phi=iw\Phi, \quad K_A\Phi=0.
\end{equation}
$S_{ab}$ are Lorentz generators in the appropriate representation, $\Delta$ and $w$ are respectively called scaling and chiral weight. Note that $K_A\Phi=0$ is a non-trivial condition that has to be carefully dealt with, for instance $\Phi$ is primary does not guarantee that $K_A\nabla_B\Phi=0$. A primary field is chiral if in addition $\bar\nabla^{\dot{\alpha}}\Phi=0$. Chiral fields have only undotted indices just like the usual supergravity case, and must have $2\Delta=3w$. For example, $W_{\alpha\beta\gamma}$ appearing in the curvature terms is a primary chiral superfield with weights $(\Delta,w)=(3/2,1)$.

General actions in conformal superspace are the familiar D-term and F-term integrations. A D-term expression is given by:
\begin{equation}\label{Dtermdef}
S_D=\int d^4xd^4\theta \, EV,
\end{equation}
$E=\det E_M{}^A$ is the (super)-determinant of the vielbein, which one can easily show
\begin{equation}\label{Eproperty}
AE=K_AE=0, \quad DE=-2E.
\end{equation}
This implies $V$ must be a real primary field with weights $(\Delta,w)=(2,0)$, and obviously it also has to be a Lorentz scalar. An F-term will be of the form
\begin{equation}\label{Ftermdef}
S_F=\int d^4xd^2\theta \, \mathcal{E} W,
\end{equation}
with the integral evaluated at the subspace $\bar\theta=0$, here $\mathcal{E}=\det{\mathcal{E}_M^A}$, $\mathcal{E}_M^A$ is $E_M{}^A$ but with dotted indices omitted. $W$ ought to be a primary chiral field with weights $(\Delta,w)=(3,2)$, again it has to be a Lorentz scalar. 

One can convert a D-term action to F-term via the \emph{chiral projector}, similar to the role of $-\frac{1}{4}(\bar{\mathcal{D}}^2-8R)$ in ordinary supergravity. The expression is surprising simple in the conformal superspace:
\begin{equation}\label{chiralprojdef}
\mathcal{P}[V]=-\frac{1}{4}\bar{\nabla}^2V.
\end{equation}
The D-term to F-term conversion formula is
\begin{equation}\label{DtoF}
\int d^4xd^4\theta\, EV=\frac{1}{2}\int d^4xd^2\theta \, \mathcal{E} \mathcal{P}[V]+\frac{1}{2}\int d^4xd^2\bar{\theta} \, \bar{\mathcal{E}} \bar{\mathcal{P}}[V].
\end{equation}

Suppose we have multiple chiral superfields $\Phi_I$, the most general action one can write down is
\begin{equation}\label{generalchiralaction}
S=[-3Z(\Phi_I,\bar{\Phi}_I)]_D+([P(\Phi_I)]_F+\textrm{h.c.}),
\end{equation}
and we can reasonably assume $Z$ is non-negative for stability. By a suitable redefinition of the fields, one can have all but one chiral field, $\Phi^i$'s, to have vanishing weights, and the remaining one, $\Phi_0$, has weights $(\Delta,w)=(1,2/3)$.\footnote{Recall primary chiral fields have fixed ratio for the weights.} This implies one can write
\begin{equation}\label{ZWdef}
Z=\bar{\Phi}_0\Phi_0\exp(-\frac{K(\Phi^i,\bar{\Phi}^{\bar{j}})}{3}), \quad P=\Phi_0^3W(\Phi^i),
\end{equation}
and we arrive at the supergravity action in the \emph{conformal compensator} formalism appearing in \cite{Kugo82}:
\begin{equation}\label{conformalSUGRAaction}
S=-3\int d^4xd^4\theta\, E\bar{\Phi}_0\Phi_0e^{-K/3}+\left(\int d^4xd^2\theta \, \mathcal{E} \Phi_0^3W+\textrm{h.c.}\right).
\end{equation}
Note that there is a natural K\"ahler structure in this model, as the action is invariant under the following K\"ahler transformation:
\begin{equation}\label{kahlertransdef}
K\rightarrow K+F+\bar{F}, \quad \Phi_0\rightarrow\Phi_0\exp(F/3), \quad W\rightarrow\exp(-F)W,
\end{equation}
$F$ is a holomorphic function of the fields $\Phi^i$. This in particular implies that the supergravity Lagragian depends only on the combination $G=K+\log|W|^2$, as in the case of Poincar\'e supergravity. 

It is easy to also introduce the Yang-Mills interaction to conformal supergravity \cite{Kugo16b}. Suppose we have an internal symmetry group with generators $\left\{X_{(r)}\right\}$, and with commutation relations:
\begin{equation}\label{YMcommutation}
[X_{(r)},X_{(s)}]=-f_{(r)(s)}{}^{(t)}X_{(t)},
\end{equation}
one just extends the super-conformal algebra to include these extra generators. We have extra gauge connection fields
\begin{equation}\label{YMguagefielddef}
W_M{}^{\mathcal{A}}X_{\mathcal{A}}
=E_M{}^AP_A+h_M{}^{\underline{\mathcal{A}}}X_{\underline{\mathcal{A}}}+\mathcal{A}_M^{(r)}X_{(r)},
\end{equation}
with the corresponding curvature being
\begin{equation}\label{YMdefromedP}
\begin{aligned}
\left[P_A,P_B\right]&=-R_{AB}{}^{\mathcal{A}}X_{\mathcal{A}} \\
&=-T_{AB}{}^CP_C-R_{AB}{}^{\underline{\mathcal{A}}}X_{\underline{\mathcal{A}}}-\mathcal{F}_{AB}^{(r)}X_{(r)}.
\end{aligned}
\end{equation}
The Yang-Mills curvature is given by the familiar expression:
\begin{equation}\label{YMcurvaturedef}
\mathcal{F}_{MN}^{(r)}=\partial_M\mathcal{A}_N^{(r)}-\partial_N\mathcal{A}_M^{(r)}-\mathcal{A}_N^{(t)}\mathcal{A}_M^{(s)}f_{(s)(t)}{}^{(r)}.
\end{equation}
We impose the constraint on this curvature:
\begin{equation}\label{YMconstraint}
\mathcal{F}_{\underline{\alpha}\underline{\beta}}^{(r)}=0,
\end{equation}
the constraint imposed guarantees that we maintain the derivative relation $\{\nabla_{\alpha},\nabla_{\beta}\}=\{\bar{\nabla}_{\dot{\alpha}},\bar{\nabla}_{\dot{\beta}}\}=0$ and $\{\nabla_{\alpha},\bar{\nabla}_{\dot{\beta}}\}=-2i\nabla_{\alpha\dot{\beta}}$. Therefore we can express the remaining curvature in terms of gaugino superfields $\mathcal{W}_{\underline{\alpha}}^{(r)}$:
\begin{equation}\label{YMguaginodef}
\begin{aligned}
&\mathcal{F}_{\alpha,\beta\dot{\beta}}^{(r)}=2i\epsilon_{\alpha\beta}\mathcal{W}_{\dot{\beta}}^{(r)}\\
&\mathcal{F}_{\dot{\alpha},\beta\dot{\beta}}^{(r)}=2i\epsilon_{\dot{\alpha}\dot{\beta}}\mathcal{W}_{\beta}^{(r)}\\
&\mathcal{F}_{\alpha\dot{\alpha},\beta\dot{\beta}}^{(r)}=-\epsilon_{\dot{\alpha}\dot{\beta}}\{\nabla_\alpha,\mathcal{W}_{\beta}^{(r)}\}-\epsilon_{\alpha\beta}\{\bar{\nabla}_{\dot{\alpha}},\mathcal{W}_{\dot{\beta}}^{(r)}\}.
\end{aligned}
\end{equation}
The gaugino $\mathcal{W}_{\alpha}^{(r)}$ is a primary chiral superfield with weights $(\Delta,w)=(3/2,1)$, and satisfies the condition
\begin{equation}\label{YMguaginoconstraint}
\nabla^\alpha\mathcal{W}_{\alpha}^{(r)}=\bar{\nabla}_{\dot{\beta}}\mathcal{W}^{(r)\dot{\beta}}.
\end{equation}
The Yang-Mills action has the usual form
\begin{equation}\label{conformalYMaction}
S_{\textrm{YM}}=\frac{1}{4}\int d^4xd^2\theta \, \mathcal{E} f_{(r)(s)}\mathcal{W}^{(r)\alpha}\mathcal{W}^{(s)}_{\alpha}+\textrm{h.c.},
\end{equation}
$f_{(r)(s)}$ being the gauge kinetic function, which is symmetric in $(r)$ and $(s)$ and transforms under gauge transformation as a symmetric product of two adjoint representations.

Combining \eqref{conformalSUGRAaction} and \eqref{conformalYMaction}, we have the \emph{conformal supergravity/matter/Yang-Mills system}. The Yang-Mills generators acts as a K\"ahler isometry on $\Phi^i$, that is, it acts like a holomorphic Killing vector field:
\begin{equation}\label{kahlerYMtransform}
X_{(r)}=V_{(r)}{}^i(\Phi)\frac{\partial}{\partial\Phi^i},
\end{equation}
the action on the conjugate field is similar. Such action is generated by a Killing potential:
\begin{equation}\label{kahlerYMtransform2}
V_{(r)}{}^iK_i=-iG_{(r)}, \quad K_i=\frac{\partial K}{\partial\Phi^i},
\end{equation}
a relation that can be inverted to obtain $V_{(r)}{}^i$.\footnote{More technical details on K\"ahler isometry can be found in \cite{BBG00}, for example.} Also, we have $\Phi_0$ and $W$ being inert under the transformation, to ensure the invariance of the Lagragian.

\subsection{Reduction to Poincar\'e Supergravity}\label{conformalSUGRA3}
We have introduced the conformal supergravity model coupled to chiral and Yang-Mills matter, it is natural to ask how it compares to the ordinary Poincar\'e supergravity. Since the conformal case is a theory with a bigger gauged isometry group, one might expect Poincar\'e supergravity to appear as a certain gauge fixed version of conformal supergravity. This is indeed the case and we shall describe below how this is achieved.

One key observation is that the gauge field $B_M$ corresponding to $D$ is rotated by the $K_A$ transformation, it is actually possible that by a suitable choice of transformation parameter, to obtain the \emph{$K_A$-guage} condition:
\begin{equation}\label{Kgaugecondition}
B_M=0.
\end{equation}
One immediate consequence will be that $D$ drops out from the covariant derivative. Since we fix the $K_A$ gauge and thus it is no longer a symmetry, its guage field $f_M{}^A$ becomes an auxiliary field instead. In particular, if we consider the constraint for the $D$-curvature, $H_{\underline{\alpha}\underline{\beta}}=0$, under this gauge choice, we have 
\begin{equation}\label{RGdef}
f_{\alpha\beta}=-\epsilon_{\alpha\beta}\bar{R}, \quad f_{\dot{\alpha}\dot{\beta}}=\epsilon_{\dot{\alpha}\dot{\beta}}R, \quad f_{\alpha\dot{\beta}}=-f_{\dot{\beta}\alpha}=-\frac{1}{2}G_{\alpha\dot{\beta}}.
\end{equation}
These fields turn out to be exactly the auxiliary fields appearing in ordinary supergravity. We also redefine the curvatures such that no $f_M{}^A$ dependence appears:
\begin{equation}\label{redefinecurve}
\tilde{R}_{AB}{}^{\mathcal{A}}=R_{AB}{}^{\mathcal{A}}+R(f)_{AB}{}^{\mathcal{A}},
\end{equation}
$\tilde{R}_{AB}{}^{\mathcal{A}}$ denotes the curvature before gauge fixing, and $R(f)_{AB}{}^{\mathcal{A}}$ means the terms in the original curvature that contain $f_M{}^A$. The gauge-fixed curvature $R_{AB}{}^{\mathcal{A}}$ will have a new set of constraints, and can be expressed in terms of the auxiliary $f_M{}^A$. The details can be found in \cite{Butter10}, the net result being that all these curvature relations are exactly those of \emph{$U(1)$-superspace} introduced in \cite{BBG00}. Therefore the gauge choice \eqref{Kgaugecondition} reduces the conformal superspace to the $U(1)$-superspace. 

The next step is to map the conformal supergravity action to the one in $U(1)_K$ supergravity. Recall the supergravity action in equation \eqref{conformalSUGRAaction}, one gets the correct normalization of the Einstein-Hilbert term by imposing:
\begin{equation}\label{DAgauge}
\Phi_0=\bar{\Phi}_0=\exp(\frac{K}{6}).
\end{equation}
This is actually possible by choosing the appropriate $D$, $A$-gauge. Since a gauge for $A$ is chosen, its gauge field is now auxiliary. The conformal chiral condition for $\Phi_0$: $\bar{\nabla}_{\dot\alpha}\exp(\frac{K}{6})=0$ and its conjugate gives
\begin{equation}\label{gaugefixedA}
A_\alpha=\frac{i}{4}\mathcal{D}_\alpha K=\frac{i}{4}K_{i}\mathcal{D}_\alpha \Phi^i, \quad A_{\dot{\alpha}}=-\frac{i}{4}\mathcal{D}_{\dot{\alpha}}K=-\frac{i}{4}K_{\bar{j}}\mathcal{D}_{\dot{\alpha}}\bar{\Phi}^{\bar{j}}.
\end{equation}
Here $\mathcal{D}_a$ is the $U(1)$ superspace covariant derivative, with only Lorentz, Yang-Mills and $A$ generators appearing, and $K_{i}=\partial_{\Phi^i}K$. After some algebra with constraints one also get:
\begin{equation}\label{gaugefixedA2}
A_{\alpha\dot{\alpha}}=\frac{i}{4}\left(K_i\mathcal{D}_{\alpha\dot{\alpha}}\Phi^i-K_{\bar{j}}\mathcal{D}_{\alpha\dot{\alpha}}\bar{\Phi}^{\bar{j}}\right)+\frac{1}{4}K_{i\bar{j}}\mathcal{D}_{\alpha}\Phi^i\bar{\mathcal{D}}_{\dot{\alpha}}\bar{\Phi}^{\bar{j}}-\frac{3}{2}G_{\alpha\dot{\alpha}},
\end{equation}
$K_{i\bar{j}}$ is the K\"ahler metric: $K_{i\bar{j}}=\partial_{\Phi^i}\partial_{\bar{\Phi}^{\bar{j}}}K$. These expression are in perfect agreement with those in the $U(1)$ superspace. Hence we have shown the equivalence between the gauge fixed conformal supergravity and $U(1)_K$ supergravity. 

Of course one can also consider a different gauge choice, a reasonable one being $\Phi_0=1$. This will give another commonly seen supergravity action,  albeit with a non-canonical Einstein-Hilbert normalization. In practice, one employs a specific rescaling of the fields such that the normalized Einstein-Hilbert term is recovered. In our language, such rescaling is nothing but a $D$ and $A$ gauge transformation that change the gauge condition to the one in \eqref{DAgauge}.

\section{Quantization of Super Yang-Mills Theory in Conformal Supergravity}\label{SUGRASYM}
Supersymmetric Yang-Mills theory in conformal supergravity will be our main object to examine, which turns out to contain both crucial ingredients of an $N=1$ supersymmetric theory: the quanta of super-Yang-Mills theory itself is a vector superfield, and chiral fields appear from the gauge fixing procedure. In the following we shall consider its quantization, in preparation for studying the one-loop effective action.

\subsection{Prepotential and Background Field Method}\label{SUGRASYM1}
In conformal supergravity, with Yang-Mills interaction, we have the covariant derivative algebra: 
\begin{equation}\label{YMcovalgebra}
\begin{aligned}
&\{\nabla_{\alpha},\nabla_{\beta}\}=\{\bar{\nabla}_{\dot{\alpha}},\bar{\nabla}_{\dot{\beta}}\}=0,\\
&\{\nabla_{\alpha},\bar{\nabla}_{\dot{\beta}}\}=-2i\nabla_{\alpha\dot{\beta}}, \\
&[\nabla_{\alpha},\nabla_{{\beta}\dot{\beta}}]=-2i\epsilon_{\alpha\beta}\mathcal{W}_{\dot{\beta}}, \quad [\nabla_{\dot{\alpha}},\nabla_{{\beta}\dot{\beta}}]=-2i\epsilon_{\dot{\alpha}\dot{\beta}}\mathcal{W}_{\beta}.
\end{aligned}
\end{equation}
This set of equations has identical form as the SYM theory in \emph{flat} superspace, except that the gaugino $\mathcal{W}_{\alpha}$ has components from the conformal algebra, apart from the usual gauge ones. Hence with some modification from the flat case, it is not difficult to handle the SYM theory in conformal superspace.

The above derivative algebra \eqref{YMcovalgebra} imposes constraints that can be solved by introducing the \emph{Yang-Mills prepotential}.\footnote{Some discussion on this can be find in, for example, \cite{BBG00}.} The spinor covariant derivatives are given by
\begin{equation}\label{defprepo}
\nabla_{\alpha}=S^{-1}\nabla_{0\alpha}S, \quad \nabla_{\dot{\alpha}}=T^{-1}\nabla_{0\dot{\alpha}}T.
\end{equation}
Here $\nabla_{0}$ with $0$ subscript denotes the covariant derivatives \emph{without} Yang-Mills interactions. These prepotentials possess the following transformation freedom:
\begin{equation}\label{prepotrans}
S\rightarrow PSg^{-1}, \quad T\rightarrow QTg^{-1},
\end{equation}
with $g$ the gauge transformation parameter, and $P$, $Q$ are respectively $\nabla_{0}$ anti-chiral and chiral.

In literature for the flat scenario, one may make use of the above freedom to arrive at the chiral representation, in which the dotted spinor derivative remains unchanged when the Yang-Mills interaction is turned on: $\nabla_{\dot{\alpha}}=\nabla_{0\dot{\alpha}}$. In this setup, one needs only to consider the chiral part, but the setback is that Hermiticity is lost: the covariant derivatives are no longer Hermitian. Alternatively, we shall consider the picture in which the chiral and anti-chiral parts are on equal footing, and Hermiticity is manifest.

Define the gauge invariant object $U=ST^{-1}$, which transform under chiral transformation as 
\begin{equation}\label{Utrans}
U\rightarrow PUQ^{-1}.
\end{equation}
The Hermitian representation is the one such that
\begin{equation}\label{Hermgauge}
\nabla_{\alpha}=U^{-1/2}\nabla_{0\alpha}U^{1/2},  \quad \nabla_{\dot{\alpha}}=U^{1/2}\nabla_{0\dot{\alpha}}U^{-1/2}.
\end{equation}
This can be achieved by choosing an appropriate $g$. This $U$ is connected to the more familiar gauge vector superfield:
\begin{equation}\label{Vdef}
U=\exp(-2iV),
\end{equation}
the $-i$ factor is from the convention that the gauge generators are chosen to be anti-Hermitian. Note that $V$ is Hermitian and the Yang-Mills gauge connection and gaugino can be expressed in terms of $V$ as in the literature, for instance \cite{WessBagger92}.

In the following, we shall employ the background field method, which has the advantage that our background expansion has the desired invariance throughout. The treatment in conformal supergravity is very similar to the flat case \cite{Grisaru79}\cite{Gates83}, which we closely follow. The background-quantum splitting we are using will be of the form:
\begin{equation}\label{defbacksplit}
S=S_BS_Q, \quad T=T_BT_Q.
\end{equation}
The transformation law for $S$ and $T$ in \eqref{prepotrans} can be interpreted in two ways, one can view it as a background transformation:
\begin{equation}\label{backsplitbtrans}
S_B\rightarrow PS_Bg^{-1}, \quad S_Q\rightarrow gS_Qg^{-1}, \quad T_B\rightarrow QT_Bg^{-1}, \quad T_Q\rightarrow gT_Qg^{-1},
\end{equation}
one can see the quantum prepotentials transform covariantly under $g$. The same transformation can be alternatively treated as a quantum transformation:
\begin{equation}\label{backsplitqtrans}
S_B\rightarrow S_B, \quad T_B\rightarrow T_B, \quad S_Q\rightarrow P_QS_Qg^{-1}, \quad T_Q\rightarrow Q_QT_Qg^{-1},
\end{equation}
where $P_Q=S_B^{-1}PS_B$, $Q_Q=T_B^{-1}QT_B$ are respectively background anti-chiral and chiral, that is, with respect to the background derivatives $\nabla_B$.

We might, similarly to the pre-split case, define $U_Q=S_QT_Q^{-1}$, which transforms as
\begin{equation}\label{Uqtrans}
U_Q\rightarrow P_QU_QQ_Q^{-1}.
\end{equation}
Again by choosing a suitable $g$, we may go between different quantum representations for the covariant derivatives. One of the most common and useful representation will be the quantum-chiral representation, in which $T_Q=1$, thus
\begin{equation}\label{qchigauge}
\nabla_{\alpha}=U_Q^{-1}\nabla_{B\alpha}U_Q,  \quad \nabla_{\dot{\alpha}}=\nabla_{B\dot{\alpha}}.
\end{equation}
The most notable feature being that the dotted derivative is identical to the background one. Another important one will be the quantum-Hermitian representation:
\begin{equation}\label{qHermgauge}
\nabla_{\alpha}=U_Q^{-1/2}\nabla_{B\alpha}U_Q^{1/2},  \quad \nabla_{\dot{\alpha}}=U_Q^{1/2}\nabla_{B\dot{\alpha}}U_Q^{-1/2},
\end{equation}
here the chiral part and the anti-chiral part are treated equally.

The quanta for the SYM is the vector superfield $V_Q$, given by 
\begin{equation}\label{Vqdef}
U_Q=\exp(-2iV_Q).
\end{equation}
It is a Hermitian, if we are in background-Hermitian representation, conformally primary superfield with vanishing weights and as usual transforms as the adjoint representation of the Yang-Mills gauge group. $V_Q$ will be the only quantum object we will care about here, hence from now on for notational simplicity, we shall, unless otherwise specified, drop the subscript distinguishing background and quantum, $V$ is always $V_Q$ and other objects are always the background ones.

\subsection{Quantum Field Action}\label{SUGRASYM2}
We are ready to find the quantum action with the background field splitting discussed. We shall consider first the case with the simplest gauge coupling, the Yang-Mills action being:
\begin{equation}\label{simpleYMaction}
S_{\textrm{YM}}=\frac{1}{4g^2}\int d^4xd^2\theta \, \mathcal{E} \tr\left(\mathcal{W}_{\textrm{YM}}^{\alpha}\mathcal{W}^{}_{\textrm{YM}\alpha}\right)+\textrm{h.c.},
\end{equation}
which corresponds to a canonical gauge kinetic function $f_{(r)(s)}=g^{-2}\delta_{(r)(s)}$. The case with a more general kinetic function will not be discussed here, as it turns out that the quantization procedure is very similar. 

For quantization and one-loop effective action, it is sufficient to expand this action up to second order in $V$. One useful observation is that the final result must be representation independent, in particular one finds the quantum-chiral representation is the most convenient, which we shall adopt. Note that the derivative algebra \eqref{YMcovalgebra} tells us that we can express the gaugino field in terms of covariant derivatives:
\begin{equation}\label{gauginoincov}
\mathcal{W}_\alpha=\frac{1}{8}[\nabla_{\dot{\alpha}},\{\nabla^{\dot{\alpha}},\nabla_{\alpha}\}].
\end{equation}
In the quantum-chiral representation, the dotted derivatives have no $V_Q$ dependence, and using \eqref{qchigauge} and \eqref{Vqdef} one easily finds that the quantum SYM gaugino is given by:
\begin{equation}\label{qYMgaugino}
\mathcal{W}_{\textrm{YM},Q\alpha}=\mathcal{W}_{\textrm{YM},B\alpha}-\frac{i}{4}\bar{\nabla}^2\nabla_{\alpha}V+\frac{1}{4}\bar{\nabla}^2[V,\nabla_{\alpha}V]+O(V^3).
\end{equation}
One can actually go further and obtain a closed form expression:
\begin{equation}\label{qYMgaugino2}
\begin{aligned}
\mathcal{W}_{\textrm{YM},Q\alpha} & =\mathcal{W}_{\textrm{YM},B\alpha}-\frac{i}{4}\bar{\nabla}^2\left[\sum_{n=0}^{\infty}\frac{(\mathcal{L}_{2iV})^n}{(n+1)!}\nabla_{\alpha}V\right] \\
& =\mathcal{W}_{\textrm{YM},B\alpha}-\frac{i}{4}\bar{\nabla}^2\frac{1-\exp(\mathcal{L}_{2iV})}{\mathcal{L}_{-2iV}}\nabla_{\alpha}V,
\end{aligned}
\end{equation}
here $\mathcal{L}_{X}$ is the commutator: $\mathcal{L}_{X}Y=[X,Y]$. But in practical calculations, one can choose a Wess-Zumino type of gauge and then only terms up to second order in $V$ will be important.

Next we have to expand the SYM action, sorted by orders of $V$. The zeroth order is just the classical action \eqref{simpleYMaction}. The first order term gives us the equation of motion:
\begin{equation}\label{simpleYMEoM}
\nabla^\alpha\mathcal{W}_{\textrm{YM}\alpha}+\textrm{h.c.}=0.
\end{equation}
The Bianchi identity on $\mathcal{W}_{\textrm{YM}\alpha}$ actually implies that the two terms in \eqref{simpleYMEoM} vanishes individually. The result here is exactly the same as the flat superspace case. Now the second order term is given by:
\begin{equation}\label{simpleYM2nd}
S^{(2)}_{\textrm{YM}}=\frac{1}{16g^2}\int d^4xd^4\theta \, E\tr\left(\nabla^{\alpha}V\bar{\nabla}^{2}\nabla_{\alpha}V-4\mathcal{W}_{\textrm{YM}}^\alpha[V,\nabla_{\alpha}V]\right)+\textrm{h.c.}.
\end{equation}
Here the chiral projector $\mathcal{P}=-\bar\nabla^2/4$ is used to convert chiral integrals into the full superspace integration. To properly quantize the theory, one has to eliminate the remaining gauge degree of freedom, and this shall be considered next. 

Note that if one wants to develop Feynman rules for this theory, the above expression, after gauge fixing procedure, will give us the gauge field propagator, and the terms higher order in $V$ will become the interaction vertices. Combined with the rules for ghost fields, which we will discuss shortly, it is possible to perform amplitude calculations graphically, just like the known case in the flat limit \cite{Grisaru79}.

\subsection{Gauge fixing and Ghost Action}\label{SUGRASYM3}
Recall that the quantum vector superfield $V$ was defined from the quantum prepotential, which has the transformation law:
\begin{equation}\label{Vtrans1}
U=\exp(-2iV), \quad U\rightarrow PUQ^{-1},
\end{equation}
here the background-quantum splitting is already applied, and the subscript $Q$ is dropped for notational simplicity. These extra gauge degrees of freedom must be fixed in the quantization procedure, and in the case of conformal supergravity, this is not much different from the flat space setting \cite{Gates83}, with only minor modifications required.

The above gauge freedom has two free parameters $P$ and $Q$, which are respectively background anti-chiral and chiral, thus one can impose a chiral and an anti-chiral gauge condition. In the flat case, one sets the gauge fixing function to be
\begin{equation}\label{flatgauge}
\bar\nabla^2V-f=0, \quad \nabla^2V-\bar{f}=0.
\end{equation}
However, in conformal superspace, such a condition is not desirable because of one subtlety: while $\bar\nabla^2V$ is chiral, it is \emph{not} conformal primary. One easy way to see this is that a primary chiral field with weights $(\Delta, w)$ must have a fixed ratio for the weights: $2\Delta=3w$, however as $V$ has zero weights, $\bar\nabla^2V$ will have conformal weights being $(1, 2)$, certainly not satisfying the primary condition.

To fix this, it is necessary to introduce the \emph{compensator}, a well known object in conformal supergravity, back when it was first developed in component approaches. For our case in particular, we introduce the superfield $X$ which is primary and has conformal weights $(2,0)$. In the usual conformal supergravity model, $X$ can just be the expression of the D-term action:
\begin{equation}\label{SUGRAXdef}
X=\Phi_0\bar\Phi_0e^{-K/3}.
\end{equation}
One use of this $X$ is to construct the \emph{associated derivatives} $\mathcal{D}_A$, which map conformal primary objects to conformal primary ones, that is, $\mathcal{D}_Af$ is automatically primary if $f$ is primary, a property that usual covariant derivatives do not have. This was originally discussed by Kugo and Uehara \cite{Kugo85} in the component approach and was also considered in superspace \cite{Butter10b}. Although these derivatives are practically very useful, such machinery is not needed here and will not be discussed further.

The introduction of this new superfield allows us to fix the problem of being not conformal primary, now $\bar\nabla^2(XV)$ is a primary chiral field. Hence we shall adopt the gauge choice:
\begin{equation}\label{YMgauge1}
\bar\nabla^2(XV)-f=0, \quad \nabla^2(XV)-\bar{f}=0.
\end{equation}
Note that in practical applications, one will compare conformal supergravity to the usual Poincar\'e one by choosing a suitable conformal gauge. We will have in that case $X\rightarrow1$ and it can be shown straightforwardly that in this limit, we have
\begin{equation}\label{gaugedown1}
\bar\nabla^2(XV)\rightarrow (\bar{\mathcal{D}}^2-8R)V,
\end{equation}
a familiar chiral projection $\bar{\mathcal{D}}^2-8R$ appearing is certainly a pleasing feature.

In ordinary Yang-Mills theory without supersymmetry, a covariant quantization with proper gauge fixing requires the famous Faddeev-Popov procedure, with ghost fields having opposite statistics showing up as a result and they have non-trivial effects in loop calculations. With supersymmetry, we also have the Faddeev-Popov procedure, but in superfield language \cite{Gates83}. The details are shown in the appendix.

With the gauge condition we are using, the final gauge fixing action will be:
\begin{equation}\label{totalgfaction}
S_{\textrm{g.f.}}=\frac{1}{8g^2\xi}\tr\int d^8z \,EX^{-2}[\bar\nabla^2(XV)\nabla^2(XV)]+S_{\textrm{gh}}
\end{equation}
with $\xi$ being a constant to be determined. The ghost contribution is 
\begin{equation}\label{ghostaction}
S_{\textrm{gh}}=\tr\int d^8z \, E\left\{X(c'+\bar{c}')\mathcal{L}_{V/2}[c-\bar{c}+\coth(\mathcal{L}_{V/2})(c+\bar{c})]+X^{-2}b\bar{b}\right\}.
\end{equation} 

The field $b$ is the so-called \emph{Nielsen-Kallosh ghost}, which has the abnormal statistics. Note that as $f$, the gauge fixing functional, is primary chiral from the gauge condition, $b$ is as well. This ghost is normally absent in the regular Faddeev-Popov procedure, and its role is to ensure that additional gauge fixing terms added to the action are properly normalized. Its appearance here is due to two reasons, one being the factor $X^{-2}$ in the action, which is included to make the Lagrangian a proper $D$-term with the correct behavior under dilation. This makes the averaging factor from the smearing process not strictly a Gaussian, as opposed to the typical case. The second reason is that the gauge condition, and thus $f$, has a non-trivial background field dependence, therefore the term with $f$ alone cannot be normalized. 

The Nielson-Kallosh ghost, as with the usual ghost fields, only takes part in loop calculations and has no classical significance. In fact, in the case of flat space SYM with background field method, where the compensator $X$ is not needed at all, it has only an effect at the one-loop level. This is because without the $X^{-2}$ factor, its action is nothing but that of a free chiral field. In conformal supergravity, if $X$ is kept classical and if we go to the conformal gauge $X=1$, we have the identical behavior that $b$ is a free field. However $X$ typically has a K\"ahler potential dependence in it, applying the background field expansion of $X$ will generate couplings between $b$ and other chiral fields. But this only gives terms with at least three quantum fields and thus does not affect the one-loop effective action.

In \eqref{ghostaction}, $c$ and $c'$ are the famous Feddeev-Popov ghosts. In particular their second order action will be
\begin{equation}\label{FPaction2nd}
S^{(2)}_{\textrm{FP}}=\int d^8z \, EX\tr(c'+\bar{c}')(c+\bar{c}).
\end{equation}
Note that without the factor $X$, one can remove the pure chiral or pure anti-chiral factors $c'c$ and $\bar{c}'\bar{c}$ via chiral projections, the result being a simple action equivalent to two free chiral ghosts. Similar to the Nielsen-Kallosh ghost, as long as only the one-loop action is of concern, we might treat $X$ as classical and set $X=1$ in conformal gauge, and thus the action simplifies as described above. But in general, when $X$ cannot be ignored, one can at least treat such a term as an effective K\"ahler potential term.

Note that we have three types of chiral ghost, similar to the usual SYM case, and they are all conformal primary. The complete action combined with the original Yang-Mills term allows one to develop the full Feynman rules for graphical treatment, similar to what was historically done for the flat space SYM, for example in \cite{Grisaru79}.

\subsection{Simplifying the Second Order Action}\label{SUGRASYM4}
The previous gauge fixing procedure gives us the full quantized action, with a vector superfield as the quanta and three types of chiral ghosts. The ghost term is simple enough, in the sense that no derivatives occur, so one can easily write down the action up to any order in quantum fields. It is thus not too difficult to perform practical calculations involving ghost fields, say with a graphical method. However the vector multiplet action requires some extra work to analyze, even if we consider only the one-loop level. We would like to study its one-loop effective action, and what we shall aim at below is to rewrite the second order action in the form:
\begin{equation}\label{simplifyaction2nd}
S_{\textrm{YM}}^{(2)}=\frac{1}{2}\int d^8z\, E\tr(V\mathcal{O}V),
\end{equation}
and obtain the key object $\mathcal{O}$.

Let us start with the original second order action:
\begin{equation}\label{simpleYMfull2nd}
\begin{aligned}
S^{(2)}_{\textrm{YM}}=\frac{1}{16g^2}\int d^8z \,E & \tr\left(\nabla^{\alpha}V\bar{\nabla}^{2}\nabla_{\alpha}V+\nabla_{\dot{\alpha}}V\nabla^2\nabla^{\dot{\alpha}}V\right. \\
& \left. -4\mathcal{W}_{\textrm{YM}}^{\alpha}[V,\nabla_{\alpha}V]+4\mathcal{W}_{\textrm{YM},\dot{\alpha}}[V,\nabla^{\dot{\alpha}}V]\right)+S^{(V)}_{\textrm{g.f.}},
\end{aligned}
\end{equation}
here $S^{(V)}_{\textrm{g.f.}}$ is the gauge-fixing term containing $V$ only and without the ghosts. The tricky part is the first two terms, with four spinor derivatives appearing. Usually in the quantization procedure, we would like to have terms with two derivatives or less, since otherwise the propagator will be much more difficult to compute and to use. This is precisely where the specific gauge choice comes in. It turns out we can eliminate terms with four derivatives and we are left with those with at most two. In fact, we shall see that the final result will involve the d'Alembertian $\Box$ in the leading term, which governs the spacetime propagation of the quantum field $V$.

Let us start with the first problematic term $\nabla^{\alpha}V\bar{\nabla}^{2}\nabla_{\alpha}V$. The idea is to manipulate this so that we have the form $V\mathcal{O}V$ for some derivative operator $\mathcal{O}$, and hope that some undesired terms with too many derivatives can be eliminated by the gauge-fixing action. This can be achieved by integration by parts. However we have to be cautious when using it, as integration by parts is non-trivial in conformal supergravity, as explained in the appendix. We have 
\begin{equation}\label{YMbyparts1}
\begin{aligned}
\nabla^{\alpha}V \bar{\nabla}^{2}\nabla_{\alpha}V&=\nabla^{\alpha}(V \bar{\nabla}^2\nabla_{\alpha}V)-V\nabla^{\alpha} \bar{\nabla}^{2}\nabla_{\alpha}V\\
&\approx -f^{\alpha B}K_B(V\bar{\nabla}^{2}\nabla_{\alpha}V)-V\nabla^{\alpha} \bar{\nabla}^{2}\nabla_{\alpha}V,
\end{aligned}
\end{equation}
where $\approx$ denotes equal up to a surface term that can be integrated out under appropriate boundary conditions, which we always assume, and we shall for simplicity not distinguish $\approx$ and $=$ from now on.

To find the effect of $K_B=(K_b,S_\beta,\bar{S}^{\dot{\beta}})$, we need an important fact that will be used multiple times: \emph{If $V$ is a primary scalar field with vanishing conformal weights, $(\Delta, w)=(0,0)$ or equivalently $DV=AV=0$, then $\nabla_\alpha V$ and $\nabla_{\dot{\alpha}}V$ are also conformal primary.} To prove this, it is sufficient to show both derivatives are annihilated by $S_\beta$ and $\bar{S}_{\dot{\beta}}$, as $\{S_\beta,\bar{S}_{\dot{\beta}}\}=2i\sigma^b_{\beta\dot{\beta}}K_b$. For $\nabla_\alpha V$, $\bar{S}_{\dot{\beta}}$ anti-commutes with $\nabla_\alpha$, so automatically annihilates $\nabla_\alpha V$. We also have
\begin{equation}\label{SQcommut}
\{S_\beta,\nabla_{\alpha}\}=(2D-3iA)\epsilon_{\beta\alpha}-2M_{\beta\alpha},
\end{equation}
the right had side vanishes when acting on $V$, thus $S_\beta$ also annihilates $\nabla_\alpha V$. The case for $\nabla_{\dot{\alpha}}V$ is similar.

Now we have to find $K_B\bar{\nabla}^{2}\nabla_{\alpha}V$. For $S_\beta$ it is easy, as $S_\beta$ and $\nabla_{\dot\alpha}$ anticommute and thus $S_\beta\bar{\nabla}^{2}\nabla_{\alpha}V=\bar{\nabla}^{2}S_\beta\nabla_{\alpha}V=0$ using the fact above. Next note that for spinors, we have $\chi^\alpha\xi_\alpha=-\chi_\alpha\xi^\alpha$ and similarly for the dotted ones. We then have 
\begin{equation}\label{YMprimary1}
\begin{aligned}
\bar{S}^{\dot{\beta}}\bar{\nabla}^{2}\nabla_{\alpha}V=&\,(-\{\bar{S}^{\dot{\beta}},\nabla^{\dot{\alpha}}\}\nabla_{\dot\alpha}-\nabla_{\dot\alpha}\{\bar{S}^{\dot{\beta}},\nabla^{\dot{\alpha}}\}+\bar\nabla^2\bar{S}^{\dot{\beta}})\nabla_{\alpha}V\\
=&\,[2M^{\dot{\beta}\dot{\alpha}}-(2D+3iA)\epsilon^{\dot{\beta}\dot{\alpha}}]\nabla_{\dot{\alpha}}\nabla_{\alpha}V\\
&+\nabla_{\dot{\alpha}}[2M^{\dot{\beta}\dot{\alpha}}-(2D+3iA)\epsilon^{\dot{\beta}\dot{\alpha}}]\nabla_{\alpha}V+0\\
=&\,(2M^{\dot{\beta}\dot{\alpha}}-2\epsilon^{\dot{\beta}\dot{\alpha}})\nabla_{\dot{\alpha}}\nabla_{\alpha}V-4\nabla_{\dot{\alpha}}\epsilon^{\dot{\beta}\dot{\alpha}}\nabla_{\alpha}V\\
=&\,2M^{\dot{\beta}\dot{\alpha}}\nabla_{\dot{\alpha}}\nabla_{\alpha}V-6\nabla^{\dot{\beta}}\nabla_{\alpha}V.
\end{aligned}
\end{equation}
Recall $M^{\dot{\beta}\dot{\alpha}}=(\bar{\sigma}^{ba}\epsilon)^{\dot{\beta}\dot{\alpha}}M_{ab}$ acts on dotted indices only, and one can show 
\begin{equation}\label{YMprimary1a}
M^{\dot{\beta}\dot{\alpha}}\nabla_{\dot{\alpha}}\nabla_{\alpha}V=(\nabla^{\dot{\beta}}\delta^{\dot{\alpha}}{}_{\dot{\alpha}}+\epsilon^{\dot{\beta}\dot{\alpha}}\nabla_{\dot{\alpha}})\nabla_{\alpha}V=3\nabla^{\dot{\beta}}\nabla_{\alpha}V.
\end{equation}
This leads to $\bar{S}^{\dot{\beta}}\bar{\nabla}^{2}\nabla_{\alpha}V=0$, which further implies $K_b\bar{\nabla}^{2}\nabla_{\alpha}V=0$ via the relation $\{S_\beta,\bar{S}_{\dot{\beta}}\}=2i\sigma^b_{\beta\dot{\beta}}K_b$ and so $\bar{\nabla}^{2}\nabla_{\alpha}V$ is primary. This is actually not a big surprise, as the term $\bar{\nabla}^{2}\nabla_{\alpha}V$ is derived from the background field expansion of the Yang-Mills gaugino, which is conformal primary. To conclude,
\begin{equation}\label{YMbyparts1a}
\nabla^{\alpha}V \bar{\nabla}^{2}\nabla_{\alpha}V=-V\nabla^{\alpha} \bar{\nabla}^{2}\nabla_{\alpha}V,
\end{equation}
which is the same as the naive result, but we will see that corrections are necessary for the gauge-fixing term. Similarly We also have
\begin{equation}\label{YMbyparts2}
\nabla_{\dot{\alpha}}V\nabla^2\nabla^{\dot{\alpha}}V=-V\nabla_{\dot{\alpha}}\nabla^2\nabla^{\dot{\alpha}}V.
\end{equation}

Let us now first turn to the gauge-fixing term, as the terms with Yang-Mills gaugino are easily dealt with. Let us use cyclicity of traces to split the term into two, treating $\bar\nabla^2$ and $\nabla^2$ symmetrically:
\begin{equation}\label{gfVaction}
S^{(V)}_{\textrm{g.f.}}=\frac{1}{16g^2\xi}\tr\int d^8z \,EX^{-2}[\bar\nabla^2(XV)\nabla^2(XV)+\nabla^2(XV)\bar\nabla^2(XV)]
\end{equation}
the reason behind this will be clear soon. We expand out the derivatives and write
\begin{equation}\label{YMgfexpand1}
\begin{aligned}
Y=X^{-1}\bar\nabla^2(XV)&=(\bar\nabla^2-8R)V+2\nabla_{\dot{\alpha}}V\nabla^{\dot{\alpha}}\log{X}\\
\bar{Y}=X^{-1}\nabla^2(XV)&=(\nabla^2-8\bar{R})V+2\nabla^{\alpha}V\nabla_{\alpha}\log{X}.
\end{aligned}
\end{equation}
We have defined as in \cite{Butter10b}
\begin{equation}\label{RRbardef}
R=-\frac{1}{8X}\bar\nabla^2X, \quad \bar{R}=-\frac{1}{8X}\nabla^2X,
\end{equation}
It is not difficult to show when we choose the conformal gauge $X\rightarrow1$, these fields reduce to the auxiliary fields with the same name in Poincar\'e supergravity. Also in that case, the terms $\nabla^{\dot{\alpha}}\log{X}$ and $\nabla^{\alpha}\log{X}$ will vanish and so we get back the usual chiral projectors of the ordinary supergravity in the right hand side of \eqref{YMgfexpand1}.

Now we invoke integration by parts formula on the term 
\begin{equation}\label{YMbyparts3}
\begin{aligned}
&\bar\nabla^2V\bar{Y}\\
=\,&\nabla_{\dot{\alpha}}(\nabla^{\dot{\alpha}}V\bar{Y})-\nabla_{\dot{\alpha}}V\nabla^{\dot{\alpha}}\bar{Y}\\
=\,&-\nabla_{\dot{\alpha}}V\nabla^{\dot{\alpha}}\bar{Y}
\end{aligned}
\end{equation}
as both $\nabla^{\dot{\alpha}}V$ and $\bar{Y}$ are primary. Applying integration by parts once more we get
\begin{equation}\label{YMbyparts3a}
\begin{aligned}
&\bar\nabla^2V\bar{Y}\\
=\,&-\nabla_{\dot{\alpha}}V\nabla^{\dot{\alpha}}\bar{Y}\\
=\,&-\nabla_{\dot{\alpha}}(V\nabla^{\dot{\alpha}}\bar{Y})+V\bar\nabla^2\bar{Y}\\
=\,&f_{\dot{\alpha}}{}^BVK_B\nabla^{\dot{\alpha}}\bar{Y}+V\bar\nabla^2\bar{Y}.
\end{aligned}
\end{equation}
Finding $K_B\nabla^{\dot{\alpha}}\bar{Y}$ is just a routine application of the superconformal algebra, from $\{S_\beta,\nabla^{\dot{\alpha}}\}=0$ and $[K_b,\nabla^{\dot{\alpha}}]=i\bar\sigma_b^{\dot\alpha\beta}S_\beta$ we immediately know applying $S_\beta$ or $K_b$ will give zero. Using the anti-commutation relation $\{\bar{S}^{\dot{\beta}},\nabla^{\dot{\alpha}}\}=(2D+3iA)\epsilon^{\dot{\beta}\dot{\alpha}}-2M^{\dot{\beta}\dot{\alpha}}$ results in
\begin{equation}\label{YMprimary2}
\bar{S}^{\dot{\beta}}\nabla^{\dot{\alpha}}\bar{Y}=8\epsilon^{\dot{\beta}\dot{\alpha}}\bar{Y}.
\end{equation}
The final result is 
\begin{equation}\label{YMbyparts3b}
\begin{aligned}
&\bar\nabla^2V\bar{Y}\\
=\,&8f_{\dot{\alpha}}{}^{\dot{\alpha}}V\bar{Y}+V\bar\nabla^2\bar{Y}\\
=\,&V\bar\nabla^2\nabla^2V+8f_{\dot{\alpha}}{}^{\dot{\alpha}}V\bar{Y}\\
&+V\left(2\nabla^{\alpha}\log{X}
\bar\nabla^2\nabla_{\alpha}-8\bar{R}\bar\nabla^2+4\nabla^{\dot{\alpha}}\nabla^{\alpha}\log{X}
\nabla_{\dot{\alpha}}\nabla_{\alpha}\right.\\
&+\left.\frac{16}{3}X^{\alpha}\nabla_{\alpha}-16\nabla_{\dot\alpha}\bar{R}\nabla^{\dot{\alpha}}-8\bar\nabla^2\bar{R}\right)V.
\end{aligned}
\end{equation}
Here we defined \cite{Butter10b} 
\begin{equation}\label{Xdef}
X_\alpha=\frac{3}{8}\bar\nabla^2\nabla_{\alpha}\log{X}, \quad X^{\dot{\alpha}}=\frac{3}{8}\nabla^2\nabla^{\dot\alpha}\log{X},
\end{equation}
they reduce to, under the conformal gauge, their $U(1)$-supergravity counterparts just like the previously defined $R$ and $\bar{R}$.

Next we have the term $2\nabla_{\dot{\alpha}}V\nabla^{\dot{\alpha}}\log{X}\bar{Y}$. One integration by parts gives
\begin{equation}\label{YMbyparts4}
\begin{aligned}
&2\nabla_{\dot{\alpha}}V\nabla^{\dot{\alpha}}\log{X}\bar{Y}\\
=\,&-2f_{\dot{\alpha}}{}^BK_B(V\nabla^{\dot{\alpha}}\log{X}\bar{Y})-2V\nabla_{\dot\alpha}(\nabla^{\dot{\alpha}}\log{X}\bar{Y})\\
=\,&-8f_{\dot{\alpha}}{}^{\dot{\alpha}}V\bar{Y}-2V(\bar\nabla^2\log{X})\bar{Y}-2V\nabla_{\dot\alpha}\log{X}\nabla^{\dot{\alpha}}\bar{Y}\\
=\,&-8f_{\dot{\alpha}}{}^{\dot{\alpha}}V\bar{Y}-2V(\bar\nabla^2\log{X})(\nabla^2+2\nabla^{\alpha}\log{X}\nabla_{\alpha}-8\bar{R})V\\
&-2V\nabla_{\dot\alpha}\log{X}(\nabla^{\dot{\alpha}}\nabla^2-2\nabla^{\alpha}\log{X}\nabla^{\dot{\alpha}}\nabla_{\alpha}+2\nabla^{\dot{\alpha}}\nabla^{\alpha}\log{X}\nabla_{\alpha}\\
&-8\bar{R}\nabla^{\dot{\alpha}}-8\nabla^{\dot{\alpha}}\bar{R})V.
\end{aligned}
\end{equation}
The correction term can be found by using
\begin{equation}\label{YMprimary3}
\begin{aligned}
\bar{S}^{\dot{\beta}}(\nabla^{\dot{\alpha}}\log{X})&=X^{-1}\bar{S}^{\dot{\beta}}(\nabla^{\dot{\alpha}}X)\\
&=X^{-1}[(2D+3iA)\epsilon^{\dot{\beta}\dot{\alpha}}-2M^{\dot{\beta}\dot{\alpha}}]X\\
&=4\epsilon^{\dot{\beta}\dot{\alpha}},
\end{aligned}
\end{equation}
we see that the correction here cancels with that of $\bar\nabla^2V\bar{Y}$. Also note that $\bar\nabla^2\log{X}=-8R-\nabla_{\dot\alpha}\log{X}\nabla^{\dot{\alpha}}\log{X}$. Combining \eqref{YMbyparts3b}, \eqref{YMbyparts4} and $-8RV\bar{Y}=-8RV(\nabla^2+2\nabla^{\alpha}\log{X}\nabla_{\alpha}-8\bar{R})V$, we get
\begin{equation}\label{YMbyparts5}
\begin{aligned}
&Y\bar{Y}\\
=\,&V\bar\nabla^2\nabla^2V+2V(U^\alpha\bar\nabla^2\nabla_{\alpha}-U_{\dot{\alpha}}\nabla^{\dot{\alpha}}\nabla^2)V\\
&+V\left[(8R+2U_{\dot{\alpha}}U^{\dot{\alpha}})\nabla^2-8\bar{R}\bar\nabla^2+(4U^{\dot{\alpha}\alpha}-4U^{\dot{\alpha}}U^{\alpha})\nabla_{\dot\alpha}\nabla_{\alpha}\right]V\\
&+V\left(\frac{16}{3}X^{\alpha}+16RU^\alpha+4U_{\dot{\alpha}}U^{\dot{\alpha}}U^\alpha-4U_{\dot\alpha}U^{\dot{\alpha}\alpha}\right)\nabla_{\alpha}V\\
&+16V(\bar{R}U_{\dot{\alpha}}-\nabla_{\dot\alpha}\bar{R})\nabla^{\dot{\alpha}}V-8V(\bar\nabla^2\bar{R}+8R\bar{R}+2U_{\dot{\alpha}}U^{\dot{\alpha}}\bar{R}-2U_{\dot{\alpha}}\nabla^{\dot{\alpha}}\bar{R})V.\\
\end{aligned}
\end{equation}
We have ordered the terms by the number of derivatives involved, and used a handy notation for derivatives of $\log{X}$: $U_\alpha=\nabla_{\alpha}\log{X}$, $U_{\dot{\alpha}\alpha}=\nabla_{\dot\alpha}\nabla_\alpha\log{X}$ and so on. It is important to note that the first derivative $U_A$ will vanish in the conformal gauge, and higher order derivatives can be expressed in terms of $R$, $\bar{R}$, $X_\alpha$, $X^{\dot{\alpha}}$ and also
\begin{equation}\label{Gadef}
G_{\alpha\dot{\alpha}}=-\frac{1}{4}(U_{\alpha\dot{\alpha}}-U_{\dot{\alpha}\alpha})-\frac{1}{2}U_{\alpha}U_{\dot{\alpha}}.
\end{equation}
This $G_{\alpha\dot{\alpha}}$ shares the characteristic  with the other four objects mentioned: it coincides with the familiar $G_{\alpha\dot{\alpha}}$ in the Poincar\'e limit.

Now $\bar{Y}Y$ is just the conjugate of \eqref{YMbyparts5}, hence we arrive at the expression
\begin{equation}\label{YMbyparts6}
\begin{aligned}
&Y\bar{Y}+\bar{Y}Y\\
=\,&V(\bar\nabla^2\nabla^2+\nabla^2\bar\nabla^2)V+2V(U^\alpha\left[\bar\nabla^2,\nabla_{\alpha}\right]+U_{\dot{\alpha}}\left[\nabla^2,\nabla^{\dot{\alpha}}\right])V\\
&+V\left\{2U_{\dot{\alpha}}U^{\dot{\alpha}}\nabla^2+2U^{\alpha}U_{\alpha}\bar\nabla^2+(8G^{\alpha\dot{\alpha}}+8U^{\alpha}U^{\dot{\alpha}})[\nabla_{\dot\alpha,}\nabla_{\alpha}]\right\}V\\
&+V\left(\frac{16}{3}X^{\alpha}-16\nabla^\alpha{R}+32RU^\alpha+4U_{\dot{\alpha}}U^{\dot{\alpha}}U^\alpha-4U_{\dot\alpha}U^{\dot{\alpha}\alpha}\right)\nabla_{\alpha}V\\
&+V\left(\frac{16}{3}X_{\dot{\alpha}}-16\nabla_{\dot\alpha}\bar{R}+32\bar{R}U_{\dot{\alpha}}+4U^{\alpha}U_{\alpha}U_{\dot{\alpha}}-4U^{\alpha}U_{\alpha\dot{\alpha}}\right)\nabla^{\dot{\alpha}}V\\
&+16VU^a\nabla_aV-8V(\bar\nabla^2\bar{R}+\nabla^2R+16R\bar{R}+2U_{\dot{\alpha}}U^{\dot{\alpha}}\bar{R}+2U^{\alpha}U_{\alpha}R\\
&-2U^{\alpha}\nabla_{\alpha}R-2U_{\dot{\alpha}}\nabla^{\dot{\alpha}}\bar{R})V.
\end{aligned}
\end{equation}

To further simplify, we use the following identity, which is equation (3.27) of \cite{Kugo16a}:
\begin{equation}\label{YMkeyidentity1}
\nabla^2\bar\nabla^2+\bar\nabla^2\nabla^2-\nabla^\alpha\bar\nabla^2\nabla_{\alpha}-\nabla_{\dot\alpha}\nabla^2\nabla^{\dot{\alpha}}=16\Box+8\mathcal{W}^\alpha\nabla_{\alpha}-8\mathcal{W}_{\dot{\alpha}}\nabla^{\dot{\alpha}}.
\end{equation}
This can be proved by showing first
\begin{equation}\label{YMkeyidentity2a}
\begin{aligned}
\left[\bar\nabla^2,\nabla_{\alpha}\right]&=\nabla_{\dot{\beta}}\{\nabla^{\dot{\beta}},\nabla_{\alpha}\}-\{\nabla_{\dot{\beta}},\nabla_{\alpha}\}\nabla^{\dot{\beta}}\\
&=2i\nabla^{\dot{\beta}}\nabla_{\alpha\dot{\beta}}+2i\nabla_{\alpha\dot{\beta}}\nabla^{\dot{\beta}}\\
&=2i([\nabla^{\dot{\beta}},\nabla_{\alpha\dot{\beta}}]+\nabla_{\alpha\dot{\beta}}\nabla^{\dot{\beta}})+2i\nabla_{\alpha\dot{\beta}}\nabla^{\dot{\beta}}\\
&=4i\nabla_{\alpha\dot{\beta}}\nabla^{\dot{\beta}}+2i(-2i\delta^{\dot{\beta}}{}_{\dot{\beta}}\mathcal{W}_\alpha)\\
&=4i\nabla_{\alpha\dot{\beta}}\nabla^{\dot{\beta}}+8\mathcal{W}_\alpha,
\end{aligned}
\end{equation}
and similarly 
\begin{equation}\label{YMkeyidentity2b}
\begin{aligned}
\left[\nabla^2,\nabla^{\dot\alpha}\right]&=\nabla^{\beta}\{\nabla_{\beta},\nabla^{\dot\alpha}\}-\{\nabla^{\dot{\alpha}},\nabla^{\beta}\}\nabla_{\beta}\\
&=2i\nabla^{\beta\dot{\alpha}}\nabla_{\beta}+2i\nabla_{\beta}\nabla^{\beta\dot{\alpha}}\\
&=4i\nabla^{\beta\dot{\alpha}}\nabla_{\beta}-8\mathcal{W}^{\dot{\alpha}}.
\end{aligned}
\end{equation}
These two equations give equation (3.26) of the quoted reference:
\begin{equation}\label{YMkeyidentity3}
\begin{aligned}
&\nabla^2\bar\nabla^2=\nabla_{\dot\alpha}\nabla^2\nabla^{\dot\alpha}+8\Box-2i\nabla^{\dot\alpha\alpha}[\nabla_{\alpha},\nabla_{\dot\alpha}]-8\mathcal{W}_{\dot{\alpha}}\nabla^{\dot{\alpha}}\\
&\bar\nabla^2\nabla^2=\nabla^{\alpha}\bar\nabla^2\nabla_{\alpha}+8\Box+2i\nabla^{\dot\alpha\alpha}[\nabla_{\alpha},\nabla_{\dot\alpha}]+8\mathcal{W}^{\alpha}\nabla_{\alpha},
\end{aligned}
\end{equation}
these immediately give the desired result. Now from \eqref{YMkeyidentity1}, it is clear that we should choose $\xi=1$ in the gauge-fixing action \eqref{gfVaction}, this removes the term with four derivatives in \eqref{YMbyparts6}. Note that the total gaugino fields $\mathcal{W}$ in \eqref{YMkeyidentity1} have only $M_{ab}$, $K_A$ and Yang-Mills components, when acting on $V$ only the Yang-Mills terms survive, thus we can safely replace $\mathcal{W}^{\alpha}$ and $\mathcal{W}_{\dot{\alpha}}$ by respectively $\mathcal{W}_{\textrm{YM}}^{\alpha}$ and $\mathcal{W}_{\textrm{YM},\dot{\alpha}}$. 

Also, using \eqref{YMkeyidentity2a} and \eqref{YMkeyidentity2b} one can convert terms with three derivatives in \eqref{YMbyparts6} into a term with fewer derivatives. This shows that after gauge fixing we can remove all terms with more than two derivatives, which is what we want to achieve.

Finally, there are two more terms in the action: $-4\mathcal{W}_{\textrm{YM}}^{\alpha}[V,\nabla_{\alpha}V]$ and $4\mathcal{W}_{\textrm{YM},\dot{\alpha}}[V,\nabla^{\dot{\alpha}}V]$. Using cyclicity of traces, integration by parts and the Bianchi identity $\nabla^\alpha\mathcal{W}_{\textrm{YM},\alpha}=\nabla_{\alpha}\mathcal{W}_{\textrm{YM}}^{\dot{\alpha}}$ we obtain
\begin{equation}\label{YMbyparts7}
-4\mathcal{W}_{\textrm{YM}}^{\alpha}[V,\nabla_{\alpha}V]+4\mathcal{W}_{\textrm{YM},\dot{\alpha}}[V,\nabla^{\dot{\alpha}}V]=8V(\mathcal{W}_{\textrm{YM}}^\alpha\nabla_{\alpha}-\mathcal{W}_{\textrm{YM},\dot{\alpha}}\nabla^{\dot{\alpha}})V.
\end{equation}
Combining all terms, we finally come to the conclusion: the second order action is
\begin{equation}\label{simplifyaction2ndb}
S_{\textrm{YM}}^{(2)}=\frac{1}{2}\tr\int d^8z\, E\left(\frac{2}{g^2}\right)V\mathcal{O}_VV,
\end{equation}
here $2/g^2$ is just an irrelevant constant, and the crucial second order differential operator $\mathcal{O}_V$ is
\begin{equation}\label{YMOV}
\begin{aligned}
\mathcal{O}_V&=\Box+\frac{1}{2}G^{\alpha\dot{\alpha}}[\nabla_{\alpha},\nabla_{\dot\alpha}]+\left(\frac{X^\alpha}{3}-\nabla^{\alpha}R+\mathcal{W}_{\textrm{YM}}^{\alpha}\right)\nabla_{\alpha}\\
&+\left(\frac{X_{\dot\alpha}}{3}-\nabla_{\dot\alpha}\bar{R}-\mathcal{W}_{\textrm{YM},\dot\alpha}\right)\nabla^{\dot{\alpha}}-\frac{1}{2}\left(\bar\nabla^2\bar{R}+\nabla^2R+16R\bar{R}\right)\\
&+\frac{i}{4}U^\alpha(\nabla^{\dot{\beta}}\nabla_{\alpha\dot{\beta}}+\nabla_{\alpha\dot{\beta}}\nabla^{\dot{\beta}})+\frac{i}{4}U_{\dot{\alpha}}(\nabla^{\beta\dot{\alpha}}\nabla_{\beta}+\nabla_{\beta}\nabla^{\beta\dot{\alpha}})\\
&+\frac{1}{8}\left(U_{\dot{\alpha}}U^{\dot{\alpha}}\nabla^2+U^{\alpha}U_{\alpha}\bar\nabla^2+4U^{\alpha}U^{\dot{\alpha}}[\nabla_{\alpha},\nabla_{\dot\alpha}]\right)\\
&+\frac{1}{4}\left(8RU^\alpha+U_{\dot{\alpha}}U^{\dot{\alpha}}U^\alpha-U_{\dot\alpha}U^{\dot{\alpha}\alpha}\right)\nabla_\alpha\\
&+\frac{1}{4}\left(8\bar{R}U_{\dot{\alpha}}+U^{\alpha}U_{\alpha}U_{\dot{\alpha}}-U^{\alpha}U_{\alpha\dot{\alpha}}\right)\nabla^{\dot{\alpha}}\\
&+U^a\nabla_a+\left(U^{\alpha}\nabla_{\alpha}R+U_{\dot{\alpha}}\nabla^{\dot{\alpha}}\bar{R}-U^{\alpha}U_{\alpha}R-U_{\dot{\alpha}}U^{\dot{\alpha}}\bar{R}\right)
\end{aligned}
\end{equation}
We have divided the expression into two parts, the first two lines remain non-zero in the Poincar\'e limit, and the rest containing $U_A$ which instead vanishes. 

One might compare this result to literature with similar calculations in ordinary supergravity, for instance the case of Abelian vector multiplet as in \cite{Buchbinder86}. One can easily reproduce most of the terms there, by carefully considering how each term above reduces to its Poincar\'e counterpart, for example one can show $\bar\nabla^2\bar{R}\rightarrow(\bar{\mathcal{D}}^2-16R)\bar{R}$, instead of the naive guess $(\bar{\mathcal{D}}^2-8R)\bar{R}$. Some discrepancy arises as we are actually going down from conformal to $U(1)$-supergravity, rather than the so-called minimal supergravity. For the latter, $X_\alpha$ and $X^{\dot{\alpha}}$ do not exist and the bosonic derivative $\mathcal{D}_a$ are defined differently.

\section{Super Heat Kernel Coefficients}\label{SHeatKernel}
\subsection{Heat Kernel Method}\label{SHeatkernel1}
Consider a superfield $\Phi$, with its quantum quadratic action being
\begin{equation}\label{superaction}
S^{(2)}=\frac{1}{2}\tr\int d^8z\, E\Phi\mathcal{O}\Phi,
\end{equation}
we can quickly generalize the non-supersymmetric scenario and conclude that the one-loop effective action is given by the analogous expression
\begin{equation}\label{super1loopaction}
\Gamma_{(1)}=\frac{i}{2} \Tr_{z} \log \mathcal{O}.
\end{equation}
Here the trace is taken over the superspace $\{z=(x,\theta,\bar\theta)\}$, in other words we are taking the \emph{supertrace}. 

It is noted that numerous methods employed in the bosonic case can also be used to analyze the supersymmetric effective action, with only minimal modifications required. In particular the Schwinger's proper-time technique \cite{DeWitt64}, which is originally developed for non-supersymmetric theory, may be applied. We define the \emph{super heat kernel} $K(z,z',\tau)$, via the differential equation
\begin{equation}\label{shkde}
\left(\mathcal{O}+i\frac{\partial}{\partial \tau}\right)K(z,z';\tau)=0,
\end{equation}
with the boundary condition being
\begin{equation}\label{shkdebc}
\lim\limits_{\tau\rightarrow 0^+}K(z,z';\tau)=E^{-1}\delta^8(z-z').
\end{equation}
Equivalently $K$ is defined by the operator expression
\begin{equation}\label{defshk}
K(z,z';\tau)=e^{i\tau\mathcal{O}}E^{-1}\delta^8(z-z'),
\end{equation}
where $\mathcal{O}$ acts on the primed variable $z'$. 

Similar to the bosonic case, this $K$ encodes information about the Green's function and one-loop effective action of the theory. Integrating $K$ over $\tau$ gives us the Green's function $G(z,z')$:
\begin{equation}\label{shktoG}
G(z,z')=i\int_0^\infty d\tau \, K(z,z';\tau), \quad \mathcal{O}G(z,z')=-E^{-1}\delta^8(z-z').
\end{equation}
Also the coinidence limit $z'\rightarrow z$ is related to the effective action:
\begin{equation}\label{shkto1loop}
\Gamma_{(1)} = - \frac{i}{2}\int_0^\infty \frac{d\tau}{\tau} \, K(\tau), \quad
K(\tau)=\int d^8z \, E \, K(z,z;\tau).
\end{equation}

If $\mathcal{O}$ is a second order differential operator, which contains the d'Alembertian as the leading term, $\mathcal{O}=\Box+...$, we may expand the heat kernel into power series containing De Witt heat kernel coefficients. The starting point is the simplest case, with the superspace being flat and $\mathcal{O}=\Box$, the heat kernel of this theory is simply
\begin{equation}\label{shkfreeflat}
K(z,z';\tau)=\frac{-i}{(4\pi \tau)^2}\exp\left(i\frac{y^ay_a}{4\tau}\right)(\theta-\theta')^2(\bar\theta-\bar\theta')^2,
\end{equation}
where $y^a=(x-x')^a-i(\theta-\theta')\sigma^a\bar{\theta}'+i\theta'\sigma^a(\bar\theta-\bar\theta')$, this $y^a$ is simply the integral of the vielbein $E^a$ along the straight line connecting $z$ and $z'$.

For a more general quadratic operator $\mathcal{O}$, we can write it as 
\begin{equation}\label{Odecompsoe}
\mathcal{O}=X^{AB}\nabla_{A}\nabla_{B}+Y^A\nabla_A+Z=\mathcal{O}'+Z.
\end{equation}
Here $\mathcal{O}'$ is the part of $\mathcal{O}$ containing derivatives, so  $\mathcal{O}'$ annihilates constants. We require that $X^{ab}=\eta^{ab}$, so that the leading term is indeed the d'Alembertian. Using the covariant derivative algebra, we can, without losing generality, further assume that the tensor $X^{AB}$ is graded-symmetric, $X^{AB}=(-1)^{AB}X^{BA}$. From now on we shall always employ the implicit grading scheme and the graded-symmetric condition is simply $X^{AB}=X^{BA}$. With this $X^{AB}$, we may construct a bilinear product of two scalar functions:
\begin{equation}\label{defshkdotpro}
\left<f,g\right>=X^{AB}(\nabla_{A}f)(\nabla_{B}g),
\end{equation}
the symmetry of $X^{AB}$ implies this product is actually symmetric.

Following the non-supersymmetric case, we propose that the heat kernel is of the form
\begin{equation}\label{defshkcurved}
K(z,z';\tau)=\frac{-i}{(4\pi \tau)^2}\exp\left(i\frac{\sigma}{2\tau}\right)\Delta^{1/2}F(z,z';\tau).
\end{equation}
$\sigma(z,z')$ is a two variable function symmetric in $z$ and $z'$, a supersymmetric analog of the geodesic interval between $z$ and $z'$ in superspace. It corresponds roughly to one half of the distance squared between $z$ and $z'$. However, it is well-known that the $N=1$ superspace has no natural metric defined, so strictly speaking the concept of distance makes no sense in superspace. Nevertheless one can treat $\sigma(z,z')$ as the curved space extension of the flat space object $y^ay_a/2$. The boundary conditions for $\sigma$ are such that it reduces to the appropriate flat limit, the details can be found in \cite{Buchbinder86}. $\Delta$ is another scalar function which is the supersymmetric version of the Van Vleck-Morrette determinant, which only arises if the space is curved, in particular $\Delta$ is identically 1 for a flat superspace. We also impose that $\Delta(z,z)=1$.

Now let us substitute this expression into the differential equation \eqref{shkde}, we get
\begin{equation}\label{sFeq}
\begin{aligned}
&\frac{1}{4\tau^2}\left(2\sigma-\left<\sigma,\sigma\right>\right)\Delta^{1/2}F+\frac{i}{2\tau}\left(\mathcal{O}'\sigma+\left<\sigma,\log\Delta\right>-4\right)\Delta^{1/2}F\\
&+\frac{i}{\tau}\Delta^{1/2}\left<\sigma,F\right>+i\Delta^{1/2}\frac{\partial}{\partial\tau}F+\mathcal{O}\Delta^{1/2}F=0.
\end{aligned}
\end{equation}
If we require $F$ to be analytic, then the $1/\tau^2$ term must be identically zero. This implies $\sigma$ must satisfy 
\begin{equation}\label{ssigmade}
\left<\sigma,\sigma\right>=2\sigma.
\end{equation}
There is a further simplification if we demand
\begin{equation}\label{svmdetde}
\mathcal{O}'\sigma+\left<\sigma,\log\Delta\right>=4.
\end{equation}
The final result is then
\begin{equation}\label{sFeq2}
\frac{1}{i}\frac{\partial}{\partial\tau}F+\frac{1}{i\tau}\left<\sigma,F\right>=\tilde{\mathcal{O}}F,
\end{equation}
where $\tilde{\mathcal{O}}$ is the operator $\tilde{\mathcal{O}}=\Delta^{-1/2}\mathcal{O}\Delta^{1/2}$. We rewrite $F$ into a power series in $\tau$:
\begin{equation}\label{sandef}
F(z,z';\tau)=\sum_{n=0}^\infty a_n\frac{(i\tau)^n}{n!}.
\end{equation}
$\{a_n\}$ are the De Witt coefficients of the super heat kernel. In terms of these coefficients, \eqref{sFeq2} becomes an iterative equation
\begin{equation} \label{saneq}
a_n+\frac{1}{n}\left<\sigma,a_n\right> =\tilde{\mathcal{O}}a_{n-1} \quad (n>0),
\end{equation}
\begin{equation} \label{saneq0}
\left<\sigma,a_0\right>=0.
\end{equation}

As in the non-supersymmetric setup, the first coefficient $a_0=\delta^2(\theta-\theta')\delta^2(\bar\theta-\bar\theta)\mathcal{I}(z,z')$ contains the \emph{parallel displacement propagator} $\mathcal{I}(z,z')$, which has the useful property \cite{Kuzenko03}
\begin{equation}\label{sa0prop}
[\nabla_{(A_1}\nabla_{A_2}\dots\nabla_{A_k)}\mathcal{I}]=0,
\end{equation}
here $[...]$ denotes the coincidence limit $z'\rightarrow z$, and $(...)$ meaning the graded symmetrization of the bracketed indices. This property theoretically allows us to obtain $[a_n]$ iteratively, by repeatedly applying \eqref{saneq} multiple times. Calculations for finding the first three coefficients of some model in this way can be found, for example, in \cite{Buchbinder86}. Similar to the original calculation by De Witt \cite{DeWitt64} for the non-supersymmetric case, this procedure quickly becomes very tedious and thus impractical beyond the first few coefficients. 

As the coincidence limit of the heat kernel is closely related to the one-loop effective action, $[a_n]$ will naturally be objects of interest. In fact analogous to the non-supersymmetric regime, the first three coefficients has the significance that they give the divergence of the theory. Using a cutoff scheme to regulate \eqref{shkto1loop},
\begin{equation}\label{s1loopreg}
\Gamma^\Lambda_{(1)} =\frac{1}{32\pi^2}\int d^8z\, E\int_{\Lambda^{-2}}^\infty \frac{d(i\tau)}{(i\tau)^3} \, F (z,z;\tau).
\end{equation}
Writing $F$ in terms of $\{a_n\}$ gives the one-loop divergence
\begin{equation}\label{s1loopdiv}
\Gamma^\Lambda_{(1)\textrm{div}} =\frac{1}{32\pi^2}\int d^8z\, E 
\left(\frac{\Lambda^4}{2}[a_0]+\Lambda^2[a_1]+\frac{1}{2}\log\Lambda^2[a_2]\right).
\end{equation}
Of course supersymmetry implies that the quartic divergence must vanish: $[a_0]=0$, thus the one-loop divergence is governed by the coincidence limit of $a_1$ and $a_2$.

\subsection{Super Heat Kernel in Conformal Supergravity}\label{SHeatKernel1.2}
In conformal supergravity, extra complications arises due to the presence of the dilation operator $D$. For example, in the case of super Yang-Mills theory, the quadratic action is
\begin{equation}\label{quadYMaction}
S^{(2)}_V=\frac{1}{2}\tr\int d^8z\, E V\mathcal{O}_VV,
\end{equation}
where $\mathcal{O}_V$ carries a non-zero $D$ charge: $[D,\mathcal{O}_V]=2\mathcal{O}_V$. Because of this non-trivial charge, it is technically not appropriate to exponentiate $\mathcal{O}_V$, as in \eqref{defshk}, to define the heat kernel. To resolve this, let us consider the quantum functional integral
\begin{equation}\label{effactionwfree}
Z[V]=\frac{\int \mathfrak{D}V\, e^{\frac{i}{2}\tr\int d^8z\, E V\mathcal{O}_VV}}{Z_{\textrm{free}}},
\end{equation}
here $Z_{\textrm{free}}$ is the functional integral for the free theory of $V$, which serves to normalize the path integral measure. In usual quantum field theory, one takes the free action to be a Gaussian:
\begin{equation}\label{usualZfree}
Z_{\textrm{free}}=\int \mathfrak{D}\phi\, e^{\frac{i}{2}\tr\int d^8z\, E \phi^2}.
\end{equation}
However this is not possible in conformal supergravity, as $V^2$ is not a valid action, lacking the correct $D$-charge. To fix the problem, we have to use the compensator $X$ and set
\begin{equation}\label{SYMZfree}
Z_{\textrm{free}}=\int \mathfrak{D}V\, e^{\frac{i}{2}\tr\int d^8z\, E XV^2}.
\end{equation}
This implies that the one-loop effective action is actually a difference of two supertraces:
\begin{equation}\label{superYM1loopaction}
\Gamma_{(1)}=\frac{i}{2} (\Tr \log \mathcal{O}_V-\Tr \log X),
\end{equation}
now this expression is perfectly $D$-invariant.

By inspecting the one-loop action above, we may now define the heat kernel of $\mathcal{O}_V$ by temporarily breaking the $D$-symmetry and choose the $D$-gauge $X=1$ so $\Tr \log X=0$.\footnote{From $DX=2X$ this is always possible by a local $D$ transformation.} In this gauge, we can forget about the $D$-charge and proceed normally, one can calculate the heat kernel of $\mathcal{O}_V$, in particular the heat kernel coefficients. To restore the $D$-symmetry, we may just insert powers of $X$ in various expressions such that we get the correct quantum number. If we are considering the one-loop action and its divergence, this procedure shall give the correct result.

Of course this is just one of the ways to regulate the $D$-symmetry, for instance one can alternatively just consider the heat kernel of $X^{-1}\mathcal{O}_V$, which is $D$-invariant. Different results may appear for different schemes, but various methods should be equivalent as long as we consider the theory on-shell.

\subsection{Non-Iterative Method for Super Heat Kernel Coefficients}\label{SHeatKernel3}
While it is in theory possible to compute heat kernel coefficients $[a_n]$ up to arbitrary order via the recursive method, the computational complexity escalates so quickly that it is not practical to do so for higher order coefficients. It would be useful to develop non-iterative techniques to effectively compute heat kernel coefficients, and we shall describe below a supersymmetric generalization of a method developed by Avramidi \cite{Avramidi91}. A short discussion of the original technique is discussed in the appendix.

There are a few restrictions that shall be imposed in order to apply such method. We shall suppose that:
\begin{enumerate}
	\item The trace of the torsion vanishes:
	\begin{equation}\label{0traceT}
	T_{AB}{}^{B}=0,
	\end{equation}
	\item For the operator $\mathcal{O}=X^{AB}\nabla_{A}\nabla_{B}+Y^A\nabla_A+Z$, we require
	\begin{equation}\label{oprestrict}
	X^{a\underline{\alpha}}=X^{\underline{\alpha}a}=Y^a=0.
	\end{equation}
\end{enumerate}
The first condition is crucial for the integration by parts formula to be true, which is very reasonable to assume, and in particular this is valid for familiar types of supergravity theory. For the second condition, it translates to the statement that the bosonic derivative $\nabla_a$ only appears in the d'Alembertian but nowhere else. This can be achieved by for example redefining the covariant derivatives, choosing certain gauges and so on. At worst we may ignore the extra terms temporarily and treat them as a perturbation later.

In the following we shall adopt a special coordinate system, the normal coordinate system in superspace $y^M=(y^m, y^\mu, y_{\dot{\mu}})$, developed in \cite{McArthur84b}. This is a straightforward supersymmetric extension of Riemann normal coordinates. We shall also choose a supersymmetric Schwinger-Fock gauge \cite{Ohrndorf86} for all the gauged symmetries, which implies that
\begin{equation}\label{sS-Fgauge}
y^Mh_M{}^{\underline{\mathcal{A}}}=0,
\end{equation}
for all gauge connections $h^{\underline{\mathcal{A}}}$, which includes the Lorentz connections, Yang-Mills connections and others. 

When defining the Schwinger heat kernel, we require that the bilinear $\sigma(z,z')$ to satisfies $\left<\sigma,\sigma\right>=2\sigma$, near the point $z$, which we shall assume to be the superspace origin from now on, $\sigma$ has the simple expression 
\begin{equation}\label{normalsigma}
\sigma=\frac{y^ay_a}{2},
\end{equation}
where $y^a=y^ME_M{}^a$. This can be shown by the properties of the normal coordinates system \cite{McArthur84b}:
\begin{equation}\label{snorcoprop}
y^ME_M{}^A=y^M\delta_M{}^A, \quad y^M\partial_M=y^M\nabla_M.
\end{equation}
For the object $\Delta$, it also simplifies in this coordinate system, which is simply:
\begin{equation}\label{normalDelta}
\Delta=E^{-1}=\det(E_A{}^M).
\end{equation}
The covariant derivative of $\Delta$ is given by:
\begin{equation}\label{DDelta}
\nabla_A\Delta=E_M{}^B\nabla_AE_B{}^M=E_M{}^{B}\nabla_BE_A{}^M-T_{AB}{}^{M}E_M{}^B=\nabla_ME_A{}^M,
\end{equation}
note that the vanishing torsion trace is used here. Using this idenitiy and with some algebra, we can show that
\begin{equation}\label{normalDeltaprove}
\begin{aligned}
&\mathcal{O}'\sigma+\left<\sigma,\log\Delta\right>\\
=&\nabla_M(X^{AB}\nabla_A\sigma E_{B}{}^M)\\
=&\partial_M(X^{aB}y_aE_{B}{}^M)=4.
\end{aligned}
\end{equation}
Be reminded that $\mathcal{O}'=X^{AB}\nabla_A\nabla_BE+Y^A\nabla_A$ is the part of $\mathcal{O}$ without the non-derivative term, and note that the constraints imposed on $\mathcal{O}$ implies $\mathcal{O}'\sigma=\Box\sigma$. Thus $\Delta$ has the desired property.

Now we have $\left<\sigma,a_n\right>=y^m\partial_ma_n$ thanks to the property \eqref{snorcoprop}. So the De Witt recursion equation becomes
\begin{equation} \label{saneqnormal}
\left(1+\frac{y^m\partial_m}{n}\right)a_n=\tilde{\mathcal{O}}a_{n-1}.
\end{equation}
To compute $a_n$, in fact the coincidence limit of the coefficients $[a_n]$, we shall expand spacetime functions with respect to the following basis, similar to the non-supersymmetric case,
\begin{equation}\label{defsket}
\begin{aligned}
&\ket{n=(a, b, c)}=\ket{a}\ket{\mu,b}\ket{\dot{\mu},c}, \quad (a\ge 0,\quad 2\ge b, c \ge 0),\\
&\ket{0}=1, \quad \ket{a}=\frac{1}{a!}y^{m_1}y^{m_2}...y^{m_a}, \quad (a>0) \\
&\ket{\mu,0}=1, \quad \ket{\mu,1}=y^\mu, \quad \ket{\mu,2}=y^\mu y_\mu,\\
&\ket{\dot{\mu},0}=1,\quad \ket{\dot{\mu},1}=y_{\dot{\mu}}, \quad \ket{\dot{\mu},2}=y_{\dot{\mu}}y^{\dot{\mu}}.
\end{aligned}
\end{equation}
Their corresponding bras are defined by
\begin{equation}\label{defsbra}
\begin{aligned}
&\bra{n}=\bra{a}\bra{\mu,b}\bra{\dot{\mu},c}, \\
&\bra{a}=\partial_{m_1}\partial_{m_2}...\partial_{m_a} \\
&\bra{\mu,0}=1, \quad \bra{\mu,1}=\partial_\mu, \quad \ket{\mu,2}=\frac{1}{4}\partial^\mu\partial_\mu,\\
&\bra{\dot{\mu},0}=1,\quad \bra{\dot{\mu},1}=\partial^{\dot{\mu}}, \quad \ket{\dot{\mu},2}=\frac{1}{4}\partial_{\dot{\mu}}\partial^{\dot{\mu}}.
\end{aligned}
\end{equation}
The inner product is given by taking the coincidence limit $y^M\rightarrow0$ after taking the derivatives, for example
\begin{equation} \label{defsbraket}
\bra{a,1,0}\ket{f}=\partial_{m_1}\partial_{m_2}...\partial_{m_a}\partial_{\mu}f|_{y^M\rightarrow 0}.
\end{equation}
The basis $\ket{n}$ is actually a complete basis, its completeness can be seen by using the supersymmetric covariant Taylor series \cite{Kuzenko03}, which is simple in normal coordinates:
\begin{equation} \label{scovtaylor}
\begin{aligned}
f(z') & =I(z',z)\sum_{n=0}^\infty \frac{1}{n!}y^{M_1}y^{M_2}\cdots y^{M_n}\nabla_{M_n}\nabla_{M_{n-1}}\cdots\nabla_{M_1}f(w)\left.\right|_{w=z}\\
&=\sum_{n=0}^\infty \frac{1}{n!}y^{M_1}y^{M_2}\cdots y^{M_n}\partial_{M_n}\partial_{M_{n-1}}\cdots\partial_{M_1}f(0),
\end{aligned}
\end{equation}
note the parallel displacement operator is just the identity, $I(z',z)=1$, in normal coordinates.

Now note that $y^m\partial_m\ket{a}=a\ket{a}$ is an eigenvector by definition, so  we have $y^m\partial_m\ket{a,b,c}=a\ket{a,b,c}$, and the zeroth-order coefficients in normal coordinates is simply $a_0=I(z,z')\delta^2(\theta-\theta')\delta^2(\bar\theta-\bar\theta')=1\times y^\mu y_\mu y_{\dot{\mu}}y^{\dot{\mu}}=\ket{0,2,2}$, by iterating 
\eqref{saneqnormal} multiple times we have:
\begin{equation} \label{sanbraket}
\begin{aligned}
\left[a_k\right]=\sum_{n_1,n_2,\dots n_{k-1}} & \left(1+\frac{a_{1}}{k}\right)^{-1}\left(1+\frac{a_2}{k-1}\right)^{-1}\cdots\left(1+a_{k-1}\right)^{-1}\\
& \times \bra{0,0,0}\tilde{\mathcal{O}}\ket{n_1}\bra{n_1}\tilde{\mathcal{O}}\ket{n_2}\cdots\bra{n_{k-1}}\tilde{\mathcal{O}}\ket{0,2,2},
\end{aligned}
\end{equation}
where the summation is over all triplet $n_i=(a_i,b_i,c_i)$. It is important that the summation is finite. This is because $\tilde{\mathcal{O}}$ is a quadratic operator and thus $\bra{a_i,b_i,c_i}\tilde{\mathcal{O}}\ket{a_{i+1},b_{i+1},c_{i+1}}$ is non-zero, after taking the coincidence limit, only if 
\begin{equation}\label{nonzerobraketcon}
a_i+b_i+c_i+2\ge a_{i+1}+b_{i+1}+c_{i+1},
\end{equation}
otherwise there are not enough derivatives to annihilate the factors of $y^M$ in $\ket{a_{i+1},b_{i+1},c_{i+1}}$.

\subsection{First Three Coefficients and Algorithm to Compute Higher Order Coefficients}\label{SHeatKernel4}
Let us compute the first few coefficients, the first coefficient is zero:
\begin{equation}\label{sa0=0}
[a_0]=\bra{0,0,0}\ket{0,2,2}=0,
\end{equation}
as expected by supersymmetry. $[a_1]$ is also trivial:
\begin{equation}\label{sa1=0}
[a_1]=\bra{0,0,0}\tilde{\mathcal{O}}\ket{0,2,2}=0,
\end{equation}
from the condition \eqref{nonzerobraketcon}. 

To compute $[a_2]$, note that the imposed constrains and graded symmetry of $X^{AB}$ imply
\begin{equation}\label{Oquadpart}
\mathcal{O}=\Box+A\nabla^2+B\bar\nabla^2+V^{\alpha\dot{\alpha}}[\nabla_\alpha,\nabla_{\dot{\alpha}}]+...,
\end{equation}
with $V^{\alpha\dot{\alpha}}=(\bar\sigma^{a})^{\alpha\dot{\alpha}}V_a$ for some bosonic vector $V_a$. It is clear that $\tilde{\mathcal{O}}=\Delta^{-1/2}\mathcal{O}\Delta^{1/2}$ and $\mathcal{O}$ share the same quadratic part. Now using \eqref{sanbraket},
\begin{equation}\label{sa2braket}
\left[a_2\right]=\sum_{n}\left(1+a\right)^{-1}\bra{0,0,0}\tilde{\mathcal{O}}\ket{n}\bra{n}\tilde{\mathcal{O}}\ket{0,2,2},
\end{equation}
there are only a few choices of $n$ to have a non-zero product, direct inspection shows that
\begin{equation}\label{sa2result}
\begin{aligned}
\left[a_2\right]&=\bra{0,0,0}\tilde{\mathcal{O}}\ket{0,1,1}\bra{0,1,1}\tilde{\mathcal{O}}\ket{0,2,2}\\
&+\bra{0,0,0}\tilde{\mathcal{O}}\ket{0,2,0}\bra{0,2,0}\tilde{\mathcal{O}}\ket{0,2,2}\\
&+\bra{0,0,0}\tilde{\mathcal{O}}\ket{0,0,2}\bra{0,0,2}\tilde{\mathcal{O}}\ket{0,2,2}\\
&=16V^{\alpha\dot{\alpha}}V_{\alpha\dot{\alpha}}+16AB+16BA=-32V^aV_a+32AB.
\end{aligned}
\end{equation}

We shall outline schematically how to obtain all the non-zero terms in the summation of \eqref{sanbraket} that contribute to $[a_k]$. By examining the structure of \eqref{sanbraket}, each non-zero product of brackets is characterized by a chain of triplets:
\begin{equation}\label{chaindef}
n_0=(0,0,0)\rightarrow n_1=(a_1,b_1,c_1) \rightarrow n_2=(a_2,b_2,c_2)\rightarrow... \rightarrow n_k=(0,2,2),
\end{equation}
with $a_i\ge 0$ and $2\ge b_i, c_i \ge 0$. Denote $s_i=a_i+b_i+c_i$, $\Delta s_i=s_i-s_{i-1}$, the chain will have to satisfy additional properties:
\begin{equation}\label{chainprop}
\begin{aligned}
\Delta s_i\le 2, \quad \Delta a_i\le 2, \quad \Delta b_i+\Delta c_i \le 2
\end{aligned}
\end{equation}
Also, if $\Delta b_i>0$ or $\Delta c_i>0$, then $\Delta a_i\le0$.\footnote{This is from the fact that the quadratic part of $\tilde{\mathcal{O}}$ satisfies $X^{a\underline{\alpha}}=X^{\underline{\alpha}a}=0$.} To find all these chains, first obtain the chain of $s_i$: $0\rightarrow s_1\rightarrow...\rightarrow s_k=4$, such that $\Delta s_i\le 2$. Then from $s_i$, obtain all pairs $b_i$ and $c_i$ with $\Delta b_i+\Delta c_i \le 2$, then check that $\Delta a_i\le 2$ and $\Delta a_i\le0$ if $\Delta b_i>0$ or $\Delta c_i>0$. This is effectively just a combinatorial problem.

Let us consider $[a_3]$ as an example. We need to find all chains $0\rightarrow s_1\rightarrow s_2 \rightarrow 4$, such that each step increases by at most two. There are only a handful of possibilities:
\begin{equation}\label{a3chain1}
\begin{aligned}
&0\rightarrow 0\rightarrow 2 \rightarrow 4,\\
&0\rightarrow 1\rightarrow 2 \rightarrow 4,\\
&0\rightarrow 1\rightarrow 3 \rightarrow 4,\\
&0\rightarrow 2\rightarrow 2 \rightarrow 4,\\
&0\rightarrow 2\rightarrow 3 \rightarrow 4,\\
&0\rightarrow 2\rightarrow 4 \rightarrow 4.
\end{aligned}
\end{equation}
Next we have to split $s_i$ into the triplet $(a_i,b_i,c_i)$, they are listed below:
\begin{enumerate}
	\item $0\rightarrow 0\rightarrow 2 \rightarrow 4$ branch:
	\begin{enumerate}
		\item $(0,0,0)\rightarrow (0,0,0)\rightarrow (0,b,c) \rightarrow (0,2,2)$ with $(b,c)$ one of $(2,0)$, $(0,2)$ or $(1,1)$.
	\end{enumerate}
	\item $0\rightarrow 1\rightarrow 2 \rightarrow 4$ branch:
	\begin{enumerate}
		\item $(0,0,0)\rightarrow (a,b,c)\rightarrow (0,b',c') \rightarrow (0,2,2)$ with $(a,b,c)$ one of $(1,0,0)$, $(0,1,0)$ or $(0,0,1)$, $(b',c')$ one of $(2,0)$, $(0,2)$ or $(1,1)$.
	\end{enumerate}
	\item $0\rightarrow 1\rightarrow 3 \rightarrow 4$ branch:
	\begin{enumerate}
		\item $(0,0,0)\rightarrow (0,b,c)\rightarrow (0,b',c') \rightarrow (0,2,2)$ with $(b,c)$ either $(1,0)$ or $(0,1)$, $(b',c')$ either $(2,1)$ or $(1,2)$.
		\item $(0,0,0)\rightarrow (1,0,0)\rightarrow (1,b',c') \rightarrow (0,2,2)$ with $(b',c')$ either $(2,0)$, $(0,2)$ or $(1,1)$ .
	\end{enumerate}
	\item $0\rightarrow 2\rightarrow 2 \rightarrow 4$ branch:
	\begin{enumerate}
		\item $(0,0,0)\rightarrow (0,b,c)\rightarrow (0,b',c') \rightarrow (0,2,2)$ with $(b,c)$ and $(b',c')$ one of $(2,0)$, $(0,2)$ or $(1,1)$.
		\item $(0,0,0)\rightarrow (2,0,0)\rightarrow (0,b',c') \rightarrow (0,2,2)$ with $(b',c')$ either $(2,0)$, $(0,2)$ or $(1,1)$ .
	\end{enumerate}
	\item $0\rightarrow 2\rightarrow 3 \rightarrow 4$ branch:
	\begin{enumerate}
		\item $(0,0,0)\rightarrow (0,b,c)\rightarrow (0,b',c') \rightarrow (0,2,2)$ with $(b,c)$ one of $(2,0)$, $(0,2)$ or $(1,1)$, $(b',c')$ either $(2,1)$ or $(1,2)$.
		\item $(0,0,0)\rightarrow (2,0,0)\rightarrow (1,b',c') \rightarrow (0,2,2)$ with $(b',c')$ either $(2,0)$, $(0,2)$ or $(1,1)$ .
	\end{enumerate}
	\item $0\rightarrow 2\rightarrow 4 \rightarrow 4$ branch:
	\begin{enumerate}
		\item $(0,0,0)\rightarrow (0,b,c)\rightarrow (0,2,2) \rightarrow (0,2,2)$ with $(b,c)$ one of $(2,0)$, $(0,2)$ or $(1,1)$.
		\item $(0,0,0)\rightarrow (0,b,c)\rightarrow (2,b,c) \rightarrow (0,2,2)$ with $(b,c)$ one of $(2,0)$, $(0,2)$ or $(1,1)$.
		\item $(0,0,0)\rightarrow (2,0,0)\rightarrow (2,b',c') \rightarrow (0,2,2)$ with $(b',c')$ either $(2,0)$, $(0,2)$ or $(1,1)$.
	\end{enumerate}
\end{enumerate}

After obtaining all the chains of triplet, one can immediately write down the coefficient $[a_k]$ in terms of various brackets, according to \eqref{sanbraket}. It remains to compute each of the terms $\bra{a_{i-1},b_{i-1},c_{i-1}}\tilde{\mathcal{O}}\ket{a_i,b_i,c_i}$, for the case of $\Delta s_i\ge0$, only one term of $\tilde{\mathcal{O}}$ will contribute, the one that exactly annihilates the factors of $y^M$ in $\ket{a_i,b_i,c_i}$. For instance, we have
\begin{equation}\label{braketex1}
\bra{a,b,c}\tilde{\mathcal{O}}\ket{a,b,c}=[Z'],
\end{equation}
where $Z'$ is the non-derivative part of $\tilde{\mathcal{O}}$: $\tilde{\mathcal{O}}=X^{AB}\nabla_{A}\nabla_{B}+Y'^A\nabla_A+Z'$. Another example will be 
\begin{equation}\label{braketex2}
\bra{1,0,0}\tilde{\mathcal{O}}\ket{0,2,0}=\bra{1,0,0}A\nabla^2\ket{0,2,0}=[4\partial_aA]=[4\nabla_aA],
\end{equation}
as we want to annihilate $y^\mu y_\mu$ appearing in $\ket{0,2,0}$. Note that the we can replace the partial derivative of $A$ by the covariant derivative, as all the connection fields vanish at the origin in the normal coordinate system.

In general, whenever we have a decrease of $a_i$, $b_i$ or $c_i$ in the bracket $\bra{a_{i-1},b_{i-1},c_{i-1}}\tilde{\mathcal{O}}\ket{a_i,b_i,c_i}$, some derivatives appearing in $\bra{a_{i-1},b_{i-1},c_{i-1}}$ will have to act on $\tilde{\mathcal{O}}$ instead of the ket $\ket{a_i,b_i,c_i}$. Hence we are required to compute the coincidence limit of derivatives of various objects in $\tilde{\mathcal{O}}$. The convenient way to do so is to use the normal coordinate expansion, in which one can expand the vielbein and other connections as a power series of $y^M$, with the coefficients expressed in terms of the field strength of the connections, in other words the torsion, Riemann curvature, and so on. For example, one can express the Yang-Mills connection in terms of the derivatives of the field strength \cite{Kuzenko03}\cite{Ohrndorf86}. More details for the case of ordinary supergravity can be found in for instance \cite{McArthur84b} and \cite{Grisaru98}, where the algorithm to compute the series expansion up to any other is shown, and the lowest order results for the vielbein are presented. 

Observing the derivative structure of $\bra{a_{i-1},b_{i-1},c_{i-1}}\tilde{\mathcal{O}}\ket{a_i,b_i,c_i}$ shows that the number of derivatives that acts on $\tilde{\mathcal{O}}$ is roughly $2+s_{i-1}-s_{i}=2-\Delta s_i$. Hence to compute the bracket we shall need normal coordinate expansion up to this order. For $[a_k]$, the maximum of $2-\Delta s_i$ is $2k-4$, which occurs when all but one of $\Delta s_i$ is 2. Thus we need the normal expansion of different objects up to $2k-4$ order, which is always possible. 

For example, to compute $[a_2]$ we have to calculate the terms in the normal coordinate expansion up to order $2(2)-4=0$, in other words we do not need the expansion at all, as demonstrated in \eqref{sa2result}. For $[a_3]$, $2k-4=2$ thus we need the second order expansion, which is not difficult to obtain as in \cite{McArthur84b} and \cite{Grisaru98}. Indeed, looking at the list of terms above, the bracket $\bra{2,0,0}\tilde{\mathcal{O}}\ket{0,b,c}$ in the $0\rightarrow 2\rightarrow 2 \rightarrow 4$ branch do require the second order normal expansion. This also happens in the $0\rightarrow 2\rightarrow 4 \rightarrow 4$ branch.

\section{One-loop Divergences of the Super Yang-Mills Theory}\label{SYMdiv}
\subsection{Heat Kernel Coefficients for Yang-Mills Vector Superfield}\label{SYMdiv1}
With the machinery developed, we are now ready to calculate the super heat kernel coefficients of SYM in conformal supergravity. First looking at the operator $\mathcal{O}_V$ derived previously in \eqref{YMOV}, in order to apply the non-recursive method, we have to satisfy the constraints. The vanishing trace of torsion is clear for conformal supergravity, however the second constraint is not satisfied, with the problematic terms being:
\begin{equation}\label{OVproblem}
\mathcal{O}_V=\frac{i}{4}U^\alpha(\nabla^{\dot{\beta}}\nabla_{\alpha\dot{\beta}}+\nabla_{\alpha\dot{\beta}}\nabla^{\dot{\beta}})+\frac{i}{4}U_{\dot{\alpha}}(\nabla^{\beta\dot{\alpha}}\nabla_{\beta}+\nabla_{\beta}\nabla^{\beta\dot{\alpha}})+U^a\nabla_a+\cdots.
\end{equation}

To resolve this, one can choose the gauge $\nabla_AX=0$, which implies $U_A=\nabla_A\log{X}=0$, thus all the terms above are eliminated. This choice of gauge implies we have to break the $D$ and $K_A$ symmetry, in other words we are reducing to the $U(1)$-supergravity regime. To ensure consistency, we have to make sure that the final result must be $D$ and $K_A$ invariant, that is, having the correct scaling dimension and being conformal primary.

Alternatively, one can employ the \emph{associated derivatives} \cite{Kugo85}\cite{Butter10b}, which is a set of modified covariant derivatives briefly mentioned before. The primary feature of this tool is that the $D$ and $K_A$ invariance are treated somewhat as a hidden symmetry, the effect of which is equivalent to choosing the gauge $\nabla_AX=0$. One can show that by rewriting $\mathcal{O}_V$ in terms of these derivatives, the undesired terms will be absorbed via the redefinition of derivatives. Then one can proceed normally using such new covariant derivatives. Of course, the two routes discussed will give the same result, for the sake of simplicity we shall choose the former one by temporarily breaking some symmetries, and manually checking the gauge invariance afterwards.

Since we are explicitly breaking the $D$ and $K_A$ symmetry, we have to redefine the covariant derivatives by removing the relevant terms. The new derivative will be exactly the covariant derivatives $\mathcal{D}_A$ in $U(1)$-supergravity. Such redefinition of course cannot alter the quadratic part of the operator $\mathcal{O}_V$, as the redefinition is a linear shift: $\mathcal{D}_A=\nabla_A+...$, thus the quadratic part of $\mathcal{O}_V$ after gauge fixing is:
\begin{equation}\label{U1OVquad}
\mathcal{O}_V=\Box+\frac{1}{2}G^{\alpha\dot{\alpha}}[\mathcal{D}_{\alpha},\mathcal{D}_{\dot\alpha}]+...
\end{equation}

Using the calculations of the previous section \eqref{sa2result}, we immediately arrive at the result:
\begin{equation}\label{SYMOVan}
[a_0]=[a_1]=0, \quad [a_2]=16\left(\frac{G^{\alpha\dot{\alpha}}}{2}\right)\left(\frac{G_{\alpha\dot{\alpha}}}{2}\right)+0=-8G_aG^a.
\end{equation}
Note that direct verification shows that $G_{\alpha\dot{\alpha}}=-\frac{1}{4}(U_{\alpha\dot{\alpha}}-U_{\dot{\alpha}\alpha})-\frac{1}{2}U_{\alpha}U_{\dot{\alpha}}$ is conformal primary, thus the expression of $[a_2]$ above is $K_A$ invariant. Also $DG_{\alpha\dot{\alpha}}=G_{\alpha\dot{\alpha}}$, so $[a_2]$ has the correct $D$-charge too. Combined with the fact that $\nabla_AX$ is not conformal primary, and thus cannot appear in $[a_2]$ before gauge fixing, we conclude that $[a_2]=-8G_aG^a$ holds identically in conformal supergravity, with $D$ and $K_A$ symmetries manifest.

It is interesting to note that, $[a_2]$ and thus the logarithmic divergence does not depend on the constant term of the operator $\mathcal{O}_V$. Typically in supergravity models, the Yang-Mills vector multiplet in general acquires a mass from the background field expansion of the K\"ahler potential term $\exp(K/3)$. The implication is that the mass of $V$ will not contribute to the one-loop divergence, a very special feature that most other theories, for instance a theory with only chiral superfields, do not have.

One can compare this result with the case of an abelian vector multiplet in minimal supergravity \cite{Buchbinder86}, the results indeed agree apart from a factor of two, which originates from different normalizations of heat kernel coefficients. The result is also consistent with the fact that in flat superspace, for pure SYM theory with Feddeev-Popov gauge-fixing, the Yang-Mills vector multiplet is UV finite and logarithmic divergence arises only from the ghost fields \cite{Ohrndorf86b}, thus $[a_2]=0$ in this case, as we can just set the auxiliary field $G_a=0$ in the case of flatness.

\subsection{Chiral Heat Kernel and One-loop Divergence from Ghost Fields}\label{SYMdiv2}
So far we have derived the heat kernel coefficients for the Yang-Mills vector multiplet $V$, but there are still ghost fields $c$, $c'$ and $b$. They are chiral superfields thus we can examine them using the \emph{chiral heat kernel}, which is just a chiral analog of the heat kernel in the full superspace. One can simply replace various full superspace quantities by the chiral subspace counterparts. For example, the heat kernel for chiral fields is defined as:
\begin{equation}\label{defchiralshk}
K(z,z';\tau)=e^{i\tau\mathcal{O}}\mathcal{E}^{-1}\delta^6(z-z'),
\end{equation}
with $z$ now the coordinates of the chiral subspace $\theta'=0$. Any superspace integration, for instance when taking the supertrace to get the one-loop effective action, must be replaced by an integration over the chiral subspace, in other words we are having an $F$-term action instead. Any techniques to compute the super heat kernel and its coefficients, can be easily applied to the chiral case, with some technical, but manageable, modifications in order to respect the chirality. One example is that one shall use a special set of normal coordinates for the chiral subspace \cite{McArthur84}, such that the coordinate functions $y'^m$ and $y'^\mu$ are chiral.\footnote{This is not the case for full superspace normal coordinates.} In particular, it is straightforward to generalize the algorithm discussed in the previous section to calculate the chiral heat kernel coefficients.

The one-loop divergence for chiral fields in conformal supergravity was discussed and calculated in \cite{Butter09}, the results therein will be applied here for the Yang-Mills ghost fields. Let us start with the Nielson-Kallosh ghost, with its action being:
\begin{equation}\label{NKghostaction}
S_{b}=\tr\int d^8z \, EX^{-2}b\bar{b}.
\end{equation}
This is close to a free action except there is an extra factor $X^{-2}$, to deal with this we can rewrite the action as $X^{-2}b\bar{b}=b\exp(-2\log{X})\bar{b}$, which is very similar to how super Yang-Mills theory couples to chiral fields: $\bar\phi\exp(2V)\phi$. Thus we can introduce an artificial $U(1)$-gauge symmetry, with $\log{X}$ taking the role of its "vector multiplet". The $\exp(-2\log{X})$ factor can be absorbed using a new covariant derivative, by adding the corresponding term for this $U(1)$. This extra $U(1)$ will have the "gaugino" field being\footnote{Recall that $X_\alpha=\frac{3}{8}\bar\nabla^2\nabla_{\alpha}\log{X}$.}:
\begin{equation}\label{U1Xgaugino}
\frac{1}{8}\bar\nabla^2e^{2\log{X}}\nabla_{\alpha}e^{-2\log{X}}=-\frac{2}{3}X_\alpha.
\end{equation}
Hence we can simply replace the Yang-Mills gaugino, whenever it shows up, by the shifted version: $\mathcal{W}^{\alpha}_{\textrm{YM}}\rightarrow\mathcal{W}^{\alpha}_{\textrm{YM}}-\frac{2}{3}X^\alpha$. Therefore from \cite{Butter09}, the one-loop divergence is given by:
\begin{equation}\label{b1loopdiv}
\begin{aligned}
\Gamma^b_{(1)\textrm{div}}=&\frac{\Lambda^2}{32\pi^2}[(1-4V_{\textrm{YM}})X]_D+\frac{\log{\Lambda^2}}{96\pi^2}S_{\chi}\\
&-\frac{\log{\Lambda^2}}{64\pi^2}\left(\left[\left(\mathcal{W}^{\alpha}_{\textrm{YM}}-\frac{2}{3}X^{\alpha}\right)^2+\frac{2}{3}W^{\alpha\beta\gamma}W_{\gamma\beta\alpha}\right]_F+\textrm{h.c.}\right),
\end{aligned}
\end{equation}
where 
\begin{equation}\label{schidef}
S_{\chi}=[G^aG_a+2R\bar{R}]_D+\left(\left[\frac{1}{12}X^\alpha X_\alpha+\frac{1}{2}W^{\alpha\beta\gamma}W_{\gamma\beta\alpha}\right]_F+\textrm{h.c.}\right)
\end{equation}
is a topological invariant. Note that we have an extra minus sign from the abnormal statistics of ghost fields, and $\log{\epsilon}=-\log{\Lambda^2}$.

Next we turn to the Feddeev-Popov ghost $c$ and $c'$, the relevant one-loop action being
\begin{equation}\label{FP1loopaction}
\begin{aligned}
S^{(2)}_{\textrm{FP}}&=\tr\int d^8z \, EX(c'+\bar{c}')(c+\bar{c})\\
&=\tr\int d^8z \, EX(c'\bar{c}+\bar{c}'c)
+\tr\int d^6z \, \mathcal{E}2RXc'c+\tr\int d^6\bar{z} \, \mathcal{\bar{E}}2\bar{R}X\bar{c}'\bar{c}.
\end{aligned}
\end{equation}
Here the chiral projector is used to produce the $F$-terms, by using the definition of $R$: $-\frac{1}{4}\bar\nabla^2X=2RX$. Similar to the case of Nielson-Kallosh ghost we absorb the factor of $X$ in the action by introducing by hand an extra $U(1)$-gauge sector, this time the shift of the gaugino is given by: $\mathcal{W}^{\alpha}_{\textrm{YM}}\rightarrow\mathcal{W}^{\alpha}_{\textrm{YM}}+\frac{1}{3}X^\alpha$. Now $2R$ and its conjugate can be treated as a mass term for the fields, again using the result of the reference \cite{Butter09}, we have
\begin{equation}\label{FP1loopdiv}
\begin{aligned}
\Gamma^{\textrm{FP}}_{(1)\textrm{div}}=&\frac{\Lambda^2}{16\pi^2}[(1-4V_{\textrm{YM}})X]_D+\frac{\log{\Lambda^2}}{48\pi^2}S_{\chi}-\frac{\log{\Lambda^2}}{16\pi^2}[4R\bar{R}]_D\\
&-\frac{\log{\Lambda^2}}{32\pi^2}\left(\left[\left(\mathcal{W}^{\alpha}_{\textrm{YM}}+\frac{1}{3}X^{\alpha}\right)^2+\frac{2}{3}W^{\alpha\beta\gamma}W_{\gamma\beta\alpha}\right]_F+\textrm{h.c.}\right),
\end{aligned}
\end{equation}
we have multiplied the divergence by a factor of $-2$ as there are two sets of ghosts with abnormal statistics.

\subsection{Full One-loop Divergence of Super Yang-Mills Theory}\label{SYMdiv3}
Finally combining the result for the Yang-Mills vector field, the Faddeev-Popov ghosts and the Nielson-Kallosh ghost, and taking the Yang-Mills traces, we present here the full one-loop divergence of the super Yang-Mills theory in conformal supergravity:
\begin{equation}\label{fullSYM1loopdiv}
\begin{aligned}
\Gamma^{\textrm{SYM}}_{(1)\textrm{div}}=&\frac{3\Lambda^2}{32\pi^2}[\tr(1-4V_{\textrm{YM}})X]_D+\frac{N_{G}\log{\Lambda^2}}{32\pi^2}S_{\chi}-\frac{N_G\log{\Lambda^2}}{8\pi^2}[G^aG_a+2R\bar{R}]_D\\
&-\frac{\log{\Lambda^2}}{64\pi^2}\left(\left[2\tr\left(\mathcal{W}^{\alpha}_{\textrm{YM}}+\frac{1}{3}X^{\alpha}\right)^2+\tr\left(\mathcal{W}^{\alpha}_{\textrm{YM}}-\frac{2}{3}X^{\alpha}\right)^2\right.\right.\\
&\left.\left.+2N_GW^{\alpha\beta\gamma}W_{\gamma\beta\alpha}\right]_F+\textrm{h.c.}\right),
\end{aligned}
\end{equation}
where $N_G=\tr1$ is the rank of the Yang-Mills gauge group. There are some cancellations for the logarithmic divergence by substituting the definition of $S_\chi$, we get
\begin{equation}\label{fullSYM1looplogdiv}
\begin{aligned}
\Gamma_{\textrm{log}}=&-\frac{3N_G\log{\Lambda^2}}{32\pi^2}[G^aG_a+2R\bar{R}]_D
\\
&-\frac{\log{\Lambda^2}}{64\pi^2}\left(\left[3\tr\mathcal{W}^{\alpha}_{\textrm{YM}}\mathcal{W}_{\textrm{YM},\alpha}+\frac{N_G}{2}X^\alpha X_\alpha+N_GW^{\alpha\beta\gamma}W_{\gamma\beta\alpha}\right]_F+\textrm{h.c.}\right),
\end{aligned}
\end{equation}
a somewhat surprisingly simple result. It is easily checked that this expression is the same as the calculation in the special case of SQED in minimal supergravity \cite{Buchbinder86}. Of course, the term with the trace of $\mathcal{W}_{\textrm{YM}}$ can be interpreted as a renormalization of the original Yang-Mills action, the pre-factor here is consistent with the known beta function, at least in the flat superspace case.

Let us consider the presented results in terms of component fields. For practical purposes, we shall apply the conformal gauge $X=1$, in other words we are reducing to the $U(1)$ supergravity. Let us start with the quadratic divergence in \eqref{fullSYM1loopdiv}. We have a term proportional to $[1]_D$, which corresponds to a renormalization of the supergravity multiplet action, its component form can be looked up from equation (4.5.6) of \cite{BBG00}:
\begin{equation}\label{SUGRAcompo}
\begin{aligned}
[1]_D=\int d^4x\,e& \left[-\frac{1}{2}\mathcal{R}+\frac{1}{2}\epsilon^{mnpq}(\bar\psi_m\bar\sigma_n\nabla_p\psi_q-\psi^m\sigma_n\nabla_p\bar\psi_q)\right.\\
&\left.-\frac{1}{3}M\bar{M}+\frac{1}{3}b^ab_a+\mathbf{D}_{\textrm{matter}}\right].
\end{aligned}
\end{equation}
Here $\mathcal{R}$ is the Ricci scalar, $\psi$ the gravitino, $M$ and $b^a$ are the auxiliary fields of the multiplet, and $\mathbf{D}_{\textrm{matter}}$ is the matter contribution which depends on the K\"ahler potential. Next we have the term $[V_{\textrm{YM}}]_D$, which is helicity-odd. In fact, it induces a divergent Fayet-Iliopoulos term which can be canceled by introducing a local counterterm \cite{Butter09}. Thus we will not further discuss this term here.

We turn to the logarithmic divergence \eqref{fullSYM1looplogdiv}. As mentioned previously there is a renormalization term for the SYM action and the full component expression is quite lengthly, which can be found in \cite{BBG00}. The remaining divergence is a linear combination of
\begin{equation}\label{topoin1}
\Delta_1=[G^aG_a+2R\bar{R}]_D+\left(\frac{1}{12}[X^\alpha X_\alpha]_F+\textrm{h.c.}\right)
\end{equation}
and
\begin{equation}\label{topoin2}
\Delta_2=[W^{\alpha\beta\gamma}W_{\gamma\beta\alpha}]_F+\textrm{h.c.}.
\end{equation}
These two expressions are actually related to topological invariants, namely the \emph{Gauss-Bonnet} and the \emph{Pontryagin} invariant. Such supergravity invariants are discussed in, for instance, \cite{Townsend79b} and \cite{Girardi91}. By using the general technique in section 4 of \cite{BBG00}, some calculation shows that the bosonic components of \eqref{topoin1} and \eqref{topoin2} are given by
\begin{equation}\label{topoincompo1}
\Delta_1=\int d^4x\,e \left(-\frac{1}{8}R^{mn}R_{mn}+\frac{1}{96}\mathcal{R}^2-\frac{1}{6}F^{mn}F_{mn}\right)
\end{equation}
and 
\begin{equation}\label{topoincompo2}
\Delta_2=\int d^4x\,e \left(\frac{1}{8}W^{mnpq}W_{mnpq}+\frac{1}{3}F^{mn}F_{mn}\right),
\end{equation}
where $R^{mn}$ is the Ricci tensor, $W^{mnpq}$ is the Weyl tensor and $F^{mn}$ is the chiral $U(1)$ curvature. Apart from the $U(1)_R$ field strength, the expressions do resemble the well-known curvature-squared invariants. Note that only the bosonic components are presented here as the remaining parts are less interesting in comparison, and one can be uniquely recover those by supersymmetry. For discussions involving the fermionic components, in the minimal supergravity formalism, one can consult for example \cite{Ferrara89}.

The appearance of such invariants comes anticipated. It can be shown \cite{Buchbinder86} that the one-loop trace of the energy-momentum tensor $T$, which directly measures the superconformal anomalies, is related to the super heat kernel coefficient $[a_2]$ and thus the logarithmic divergence:\footnote{A difference of a factor of 2 between here and \cite{Buchbinder86} is due to a different normalization.}
\begin{equation}\label{superconfano}
<T>=\frac{1}{64\pi^2}[a_2].
\end{equation}
By analyzing the super-Weyl cohomology \cite{Bonora85}, it can be shown that in the absence of background matter, the superconformal anomaly must be constructed from the Gauss-Bonnet and the Pontryagin invariant. Our result here is in agreement with this statement, as the original analysis was performed in minimal supergravity in which the $U(1)_R$ curvature is absent. It is worth noting that in each of the individual logarithmic divergence of the various fields, equation \eqref{SYMOVan}, \eqref{b1loopdiv} and \eqref{FP1loopdiv}, they all contain non-topological invariant terms. Only when we add up the contributions to obtain the total divergence \eqref{fullSYM1looplogdiv}, the problematic terms combine nicely into a multiple of $\Delta_1$. This provides a strong consistency check for our calculations.

\section{Conclusion}
We have derived the one-loop divergence of the super Yang-Mills theory in conformal supergravity, by first quantizing the theory with background field method and obtaining the second order action. The main tool employed is the heat kernel method, applied to superspaces. We have described a non-recursive technique to compute the heat kernel coefficients for the theory, and explicitly computed the first three coefficients. The method presented here theoretically allows us to compute the coefficients up to any order, as demonstrated.

The developed technique can be readily applied to other supersymmetric theories, and it will be interesting to apply such machinery to not just the Yang-Mills theory, but different interactions with distinct field contents. It is hoped that one can derive the one-loop divergence of various theories using such method. Also, one may consider a more general version of Yang-Mills theory, characterized by a non-trivial gauge kinetic function. Such generalization typically arises from different phenomenological models, for instance from string theory models. The study of such general class of super Yang-Mills theory will be a subject of future work.

\section*{Acknowledgments}
The author would like to thank Mary K. Gaillard in gratitude for inspiring discussions, helpful comments, and encouragement for this work. This work was supported in part by the Director, Office of Science, Office of High Energy and Nuclear Physics, Division of High Energy Physics, of the US Department of Energy under Contract DE-AC02-05CH11231 and in part by the National Science Foundation under grant PHY-1316783.

\newpage
\appendix
\appendixpage
\section{Feddeev-Popov Procedure in Superspace and Derivation of the Ghost Action}
In the following we shall derive the gauge-fixed action of the super Yang-Mills theory with ghost fields. Let us start by considering the functional integral for the field V:
\begin{equation}\label{funint1}
Z=\int \mathfrak{D}V\, e^{iS_{\textrm{YM}}}.
\end{equation}
To impose the gauge condition \eqref{YMgauge1}, one introduces a delta functional to the above path integral, which becomes
\begin{equation}\label{funint2}
Z=\int \mathfrak{D}V\, \Delta_{\textrm{FP}}^{-1}\, \delta(\bar\nabla^2(XV)-f)\,\delta(\nabla^2(XV)-\bar{f})e^{iS_{\textrm{YM}}},
\end{equation}
with $\Delta_{\textrm{FP}}$ the famous Faddeev-Popov determinant, which will be computed later. As in the usual case, one may average over $f$ and $\bar{f}$ in the gauge fixing function with certain weight. The standard one is the Gaussian smearing, but with a slight twist here for our scenario. Instead we insert into the functional integral, the factor
\begin{equation}\label{funint3}
1=\int\mathfrak{D}f\mathfrak{D}\bar{f}\mathfrak{D}b\mathfrak{D}\bar{b}\,\exp\left[i\frac{1}{8g^2\xi}\int d^8z \, EX^{-2}\tr(f\bar{f}+b\bar{b})\right]
\end{equation}
with $b$ having opposite statistics as $f$ to normalize the factor, this contributes to the action
\begin{equation}\label{gaugefixaction1}
S_{\textrm{GF}}=\frac{1}{8g^2\xi}\int d^8z \, EX^{-2}\tr(f\bar{f}+b\bar{b}).
\end{equation}

By rescaling $b$ we get the Nielson-Kallosh ghost action in \eqref{totalgfaction}. We then turn to the computation of the Faddeev-Popov determinant. To do so, it is necessary to know how the gauge fixing function \eqref{YMgauge1} changes under a gauge transformation. First let us rewrite the transformation law in \eqref{Vtrans1} by defining $P=\exp{-2i\bar\Lambda}$ and $Q=\exp{2i\Lambda}$, we have the familiar expression
\begin{equation}\label{Vtrans2}
e^{V'}=e^{\bar\Lambda}e^{V}e^{\Lambda}.
\end{equation}
It is well known that one can obtain the infinitesimal change in closed form:
\begin{equation}\label{Vtrans3}
\begin{aligned}
\delta{V} & =\mathcal{L}_{V/2}[\Lambda-\bar\Lambda+\coth(\mathcal{L}_{V/2})(\Lambda+\bar\Lambda)]\\
& =\Lambda+\bar\Lambda+O(V).
\end{aligned}
\end{equation}
This will be relevant in the ghost action.

We can now write the Faddeev-Popov determinant as
\begin{equation}\label{FPdet1}
\Delta_{\textrm{FP}}=\int\mathfrak{D}\Lambda\mathfrak{D}\bar\Lambda\, \delta(F-f)\,\delta(\bar{F}-\bar{f}),
\end{equation}
with $F$ and $\bar{F}$ the gauge functions: $F=\bar\nabla^2(XV)$, $\bar{F}=\nabla^2(XV)$. Here the path integral is over the gauge group parameter space. As in the usual covariant quantization procedure, one uses gauge invariance and properties of the delta function to rewrite the delta functionals. We then obtain
\begin{equation}\label{FPdet2}
\Delta_{\textrm{FP}}=\int\mathfrak{D}\Lambda\mathfrak{D}\bar\Lambda\, \delta\left(\left.\frac{\delta{F}}{\delta\Lambda}\right|\Lambda+\left.\frac{\delta{F}}{\delta\bar\Lambda}\right|\bar\Lambda\right)\,\delta\left(\left.\frac{\delta{\bar{F}}}{\delta\Lambda}\right|\Lambda+\left.\frac{\delta{\bar{F}}}{\delta\bar\Lambda}\right|\bar\Lambda\right),
\end{equation}
the stroke $|$ denoting evaluation at the "origin" $\Lambda=\bar\Lambda=0$. The next step is to recast the delta functionals using their integral representation, introducing the new fields $\Lambda'$ and $\bar\Lambda'$ with the obvious chirality:
\begin{equation}\label{FPdet3}
\begin{aligned}
\Delta_{\textrm{FP}}=\int\mathfrak{D}\Lambda\mathfrak{D}\bar\Lambda\mathfrak{D}\Lambda'\mathfrak{D}\bar\Lambda'\, \exp & \left[i\tr\int d^4xd^2\theta \, \mathcal{E}\Lambda'\left(\left.\frac{\delta{F}}{\delta\Lambda}\right|\Lambda+\left.\frac{\delta{F}}{\delta\bar\Lambda}\right|\bar\Lambda\right.\right) \\
& \left. +i\tr\int d^4xd^2\bar\theta\,\bar{\mathcal{E}}\bar\Lambda'\left(\left.\frac{\delta{\bar{F}}}{\delta\Lambda}\right|\Lambda+\left.\frac{\delta{\bar{F}}}{\delta\bar\Lambda}\right|\bar\Lambda\right)\right].
\end{aligned}
\end{equation}
What we desire is the reciprocal of $\Delta_{\textrm{FP}}$, as in the functional integral $Z$, this can be achieved by replacing the fields appearing in $\Delta_{\textrm{FP}}$ by ghost fields with opposite statistics, which at the end introduces the Faddeev-Popov action:
\begin{equation}\label{FPaction}
\begin{aligned}
S_{\textrm{FP}}&=\tr\int d^4xd^2\theta \, \mathcal{E}c'\left(\left.\frac{\delta{F}}{\delta{c}}\right|c+\left.\frac{\delta{F}}{\delta\bar{c}}\right|\bar{c}\right)+\tr\int d^4xd^2\bar\theta\,\bar{\mathcal{E}}\bar{c}'\left(\left.\frac{\delta{\bar{F}}}{\delta{c}}\right|c+\left.\frac{\delta{\bar{F}}}{\delta\bar{c}}\right|\bar{c}\right)\\
& =\tr\int d^4xd^2\theta \, \mathcal{E}c'\bar\nabla^2\left(X\delta{V}\right)+\tr\int d^4xd^2\bar\theta\,\bar{\mathcal{E}}\bar{c}'\nabla^2\left(X\delta{V}\right) \\
& =\int d^8z \, EX\tr(c'+\bar{c}')\mathcal{L}_{V/2}[c-\bar{c}+\coth(\mathcal{L}_{V/2})(c+\bar{c})].
\end{aligned}
\end{equation}
Here we have used the chiral projection to convert chiral integrals into $D$-terms, and rescaled the fields to remove the numerical pre-factor that appears. Apart from the compensator $X$, this is the same as in the case of flat space. 

As a remark, the Faddeev-Popov ghosts have vanishing conformal weights, and the Nielsen-Kallosh ghost has weights $(\Delta,w)=(3,2)$. Having dimension 3 may sound awkward and one might think this will pose technical difficulties. It turns out that its action is simple enough that this will not be a concern, especially for one-loop calculations.

\section{Integration by Parts in Conformal Supergravity}

In usual supergravity theory, if one has a superfield $v^A$, one can easily show that the term 
\begin{equation}\label{byparts1}
\int d^8z \, E\nabla_Av^A
\end{equation}
is a surface term that vanishes given the appropriate boundary conditions, hence one can safely treat them as zero. Here $A$ can be either a vector or spinor index, $a$, $\alpha$ or $\dot{\alpha}$. However in conformal superspace, such total derivative terms actually may not vanish, since there are extra generators, in particular the special conformal ones, that act non-trivially on $v^A$. In fact, special conformal curvature terms will appear, as we will derive here.

Let us start with the covariant derivative
\begin{equation}\label{byparts2}
\nabla_M(EE_A{}^Mv^A)=\partial_M(EE_A{}^Mv^A)-h_M{}^{\underline{\mathcal{A}}}X_{\underline{\mathcal{A}}}(EE_A{}^Mv^A).
\end{equation}
The first term of the right hand side is the surface term that can be neglected. For the second term, only the special conformal curvature $K_A$ may give a non-zero result, 
\begin{equation}\label{byparts3}
h_M{}^{\underline{\mathcal{A}}}X_{\underline{\mathcal{A}}}(EE_A{}^Mv^A)=Ef_{A}{}^BK_Bv^A,
\end{equation}
where $f_{A}{}^B$ is the special conformal gauge field. Note that as $K_A$ does not commute with the covariant derivatives, so even if $\nabla_Av^A$ is conformal primary, $K_B\nabla_Av^A$=0, it does not imply $v^A$ is also primary. This is exactly the reason that the usual integration by parts has to be modified in conformal superspace.

Now consider the left hand side, $\nabla_M$ acting on $v^A$ gives the desired term $E\nabla_Av^A$. There is one more term showing up when $\nabla_M$ acts on the vielbein, which can be computed by mimicking the calculation in ordinary supergravity, or in $U(1)$-supergravity \cite{BBG00}. The resulting expression is identically the same as the aforementioned cases, with torsion coefficients appearing: 
\begin{equation}\label{byparts4}
\nabla_M(EE_A{}^Mv^A)=ET_{AB}{}^Bv^A+E\nabla_Av^A.
\end{equation}
In conformal supergravity, one can easily check that the torsion term has vanishing trace similar to the ordinary scenario. We then arrive at the integration by parts formula:
\begin{equation}\label{byparts5}
\nabla_Av^A\approx-f_{A}{}^BK_Bv^A,
\end{equation}
with $\approx$ denoting equal up to a surface term that can effectively set to zero. 

One notational remark is that since the implicit grading is used throughout here, one may need to insert terms like $(-1)^A$ if such implicit grading is lifted, and such factors do appear in the usual supergravity integration by parts formula.

\section{Avramidi's Method for Non-supersymmetric Theories}
We shall present an efficient technique of deriving heat kernel coefficients developed by Avramidi \cite{Avramidi91}. For simplicity, we shall only consider operators of the form 
\begin{equation}\label{simpleO}
\mathcal{O}=\nabla_\mu\nabla^\mu+\mathcal{Q}(x).
\end{equation}

First it will be convenient to work in special coordinates, the \emph{Riemann normal coordinates} \cite{Poisson11}. Near the point $x'$, define a coordinate system $\{y^m\}$ which its coordinate function satisfies
\begin{equation}\label{normcodef2}
y^m=-\delta^{m}_{a}e_{\mu'}^a(x')\nabla^{\mu'}\sigma.,
\end{equation}
here we have introduced a moving frame with vielbein $e_\mu^a$. The recursion relation in this case is 
\begin{equation} \label{aneqnorm}
\left(1+\frac{\mathcal{D}}{n}\right)a_n =\mathcal{\tilde{O}}a_{n-1} \quad (n>0),
\end{equation}
\begin{equation} \label{aneqnorm0}
\mathcal{D}a_0 =0, \quad a_0(x,x)=1
\end{equation}
where 
\begin{equation} \label{defDOtilde}
\mathcal{D}=\nabla_\mu\sigma\nabla^\mu=y^m\nabla_m=y^m\partial_m, \quad \mathcal{\tilde{O}}=\Delta^{-\frac{1}{2}}\nabla_\mu\nabla^\mu\Delta^{\frac{1}{2}}+\mathcal{Q}.
\end{equation}
Note that $y^m\nabla_m=y^m\partial_m$ is a consequence of using Riemann normal coordinates. The Van Vleck-Morette determinant also simplifies in this coordinate systems:
\begin{equation} \label{vmdetnorm}
\Delta=\frac{1}{\sqrt{-g(x)}}=e(x)^{-1}.
\end{equation}
The coefficient $a_0$ is a known quantity, the so-called \emph{parallel displacement operator}:
\begin{equation} \label{Avraa0}
a_0=\mathcal{I}(x,x').
\end{equation}
$\mathcal{I}(x,x')$ parallel transports a field $\phi$ at $x'$ to the point $x$, and it is just identity for scalars. It satisfies the following key properties:
\begin{equation} \label{a0prop1}
[\mathcal{I}]=1, \quad [\nabla_{(\mu_1}\nabla_{\mu_2}\dots\nabla_{\mu_k)}\mathcal{I}]=0.
\end{equation}
Here square bracket means we are taking the coincidence limit:
\begin{equation} \label{bracketdef}
[f(x,x')](y)=f(y,y),
\end{equation}
this common convention will appear often from now on.

We are ready to evaluate the coefficients ${a_n}$ by formally solving \eqref{aneqnorm}:
\begin{equation} \label{Avraan}
a_n=\left(1+\frac{\mathcal{D}}{n}\right)^{-1}\mathcal{\tilde{O}}\left(1+\frac{\mathcal{D}}{n-1}\right)^{-1}\mathcal{\tilde{O}}\cdots\left(1+\mathcal{D}\right)^{-1}\mathcal{\tilde{O}}\mathcal{I}.
\end{equation}
To compute this, notice that 
\begin{equation} \label{defket1}
\mathcal{D}y^m=y^m
\end{equation}
implies $y^a$ is the eigenvector of $\mathcal{D}$ with eigenvalue 1. This allows us to define the eigenvector $\ket{n}$ with eigenvalue being any positive integer $n$, using symmetric products:
\begin{equation} \label{defketn}
\ket{0}=1, \quad \ket{n}=\frac{1}{n!}y^{m_1}y^{m_2}\cdots y^{m_n} \quad (n>0).
\end{equation}
The dual $\bra{n}$ will be defined via
\begin{equation} \label{defbra}
\bra{n}\ket{\phi}=\partial_{m_1}\partial_{m_2}\cdots\partial_{m_n}\phi\left.\right|_{y^a=0}.
\end{equation}
The orthonormal relation is easily seen satisfied. We also have the completeness relation:
\begin{equation} \label{braketcom}
\sum_{n=0}^\infty \ket{n}\bra{n}=1.
\end{equation}
To show this, we fix the point $x'$, and consider $x$ close to $x'$. The \emph{covariant Taylor series} \cite{Barvinsky85} for scalar function around $x'$ is
\begin{equation} \label{covtaylor}
\begin{aligned}
f(x) & =\sum_{n=0}^\infty \frac{1}{n!}\nabla^{\mu_1}\sigma \nabla^{\mu_2}\sigma \cdots \nabla^{\mu}\sigma \nabla_{\mu_n}\nabla_{\mu_2}\cdots\nabla_{\mu_n}f(z)\left.\right|_{z=x'} \\
& =\sum_{n=0}^\infty \frac{1}{n!}y^{m_1}y^{m_2}\cdots y^{m_n}\partial_{m_1}\partial_{m_2}\cdots\partial_{m_n}f(z)\left.\right|_{z=x'},
\end{aligned}
\end{equation}
which immediately gives the desired formula \eqref{braketcom}. Note that if the object being acted on is not a scalar, which will not concern us here, we have instead:
\begin{equation} \label{braketcomgen}
\mathcal{I}\sum_{n=0}^\infty \ket{n}\bra{n}=1.
\end{equation}
Also note that the coincidence limit is just given by a simple bracket:
\begin{equation} \label{coinbraket}
[\phi]=\bra{0}\ket{\phi},
\end{equation}
as $y^a=0$ is equivalent to $x=x'$.

Using the tools just introduced, the operator inverse appears in \eqref{Avraan} can be written as:
\begin{equation} \label{opinverse}
\left(1+\frac{\mathcal{D}}{k}\right)^{-1}=\sum_m\left(1+\frac{l}{k}\right)^{-1}\ket{l}\bra{l}.
\end{equation}
It is useful to note that it commutes with $\mathcal{I}$ as $\mathcal{DI}=0$. Now we have 
\begin{equation} \label{anbraket}
\begin{aligned}
\left[a_n\right]=\sum_{l_1,l_2,\dots l_{n-1}} & \left(1+\frac{l_{n-1}}{n}\right)^{-1}\left(1+\frac{l_{n-2}}{n-1}\right)^{-1}\cdots\left(1+l_1\right)^{-1}\\
& \times \bra{0}\mathcal{P}\ket{l_{n-1}}\bra{l_{n-1}}\mathcal{P}\ket{l_{n-2}}\cdots\bra{l_1}\mathcal{P}\ket{0},
\end{aligned}
\end{equation}
here we have used $\bra{0}\mathcal{I}\ket{k}=\delta_{k0}$, and 
\begin{equation} \label{Pdef}
\mathcal{P}=\mathcal{I}^{-1}\mathcal{\tilde{O}}\mathcal{I}.
\end{equation}
One important fact is that since $\mathcal{P}$ is a quadratic differential operator, $\bra{k}\mathcal{P}\ket{l}$ is non-zero only if $l\le k+2$, hence the summation is actually finite. It is convenient to decompose $\mathcal{P}$ into the following form:
\begin{equation} \label{Pdecom}
\mathcal{P}=X^{mn}\partial_m\partial_n+Y^m\partial_m+Z,
\end{equation}
sorted by the number of derivatives appeared, which for the operator as in \eqref{simpleO}, we have
\begin{equation} \label{simpleQX}
X^{mn}=g^{mn},
\end{equation}
\begin{equation} \label{simpleQY}
Y^{m}=2\phi^m,
\end{equation}
\begin{equation} \label{simpleQZ}
Z=\partial_m\phi^m+\phi_m\phi^m+\frac{1}{2}\partial_m\partial^mB-\frac{1}{4}\partial_mB\partial^mB+\mathcal{Q},
\end{equation}
where $\phi^m$ is the connection: $\nabla_m=\partial_m+\phi_m$, and 
\begin{equation} \label{defB}
B=\log{\Delta}=\log{e^{-1}}.
\end{equation}
Note that in normal coordinates, $\mathcal{I}$ is just the unity $\mathcal{I}=1$.

Let us compute the coefficients $[a_0]$, $[a_1]$ and $[a_2]$. $[a_0]$ is trivial: $[a_0]=1$. For $[a_1]$, it is simply
\begin{equation} \label{simplea1.1}
[a_1]=\bra{0}\mathcal{P}\ket{0}=[Z]=\mathcal{Q}+[\partial_m\phi^m]+[\phi_m\phi^m]+\frac{1}{2}[\partial_m\partial^mB]-\frac{1}{4}[\partial_mB][\partial^mB].
\end{equation}
To proceed, we need the expansion for the connection $\phi_m$, note that $y^m\nabla_m=y^m\partial_m$ implies
\begin{equation} \label{phiFSgauge}
y^m\phi_m=0.
\end{equation}
This is analogous to the Fock-Schwinger gauge in gauge theory, hence the connection will have an expansion \cite{Shifman80}:
\begin{equation} \label{phinormalexp}
\phi_m=\sum_{k=0}^\infty \frac{(\nabla_y)^k}{k!(k+2)}y^nF_{nm}, \quad \nabla_y=y^m\nabla_m.
\end{equation}
$F_{nm}$ is the field strength for the connection. Hence 
\begin{equation} \label{phibrakcet1}
[\phi_m\phi^m]=0, \quad [\partial_m\phi^m]=\frac{1}{2}F^m{}_m=0.
\end{equation}
We also use the vielbein expansion in normal coordinates \cite{Poisson11}:
\begin{equation} \label{simpleeexpand}
e_m^a(x)=\delta_m^a+\frac{1}{3!}y^ny^p\delta_n^bR_{pmb}{}^a(x')+\dots
\end{equation}
and the formula for determinant:
\begin{equation} \label{simpledetexpand}
B=\log\det (e_m^a)^{-1}=-\tr\log{e_m^a}
=-\sum_{k=1}^\infty\frac{(-1)^{k+1}}{k}\tr(e_m^a-\delta_m^a)^k.
\end{equation}
We see that 
\begin{equation} \label{Bbraket1}
[\partial_mB]=0, \quad [\partial_m\partial^mB]=\frac{1}{3}R.
\end{equation}
with $R$ the Ricci scalar. Thus
\begin{equation} \label{simplea1.2}
[a_1]=\mathcal{Q}+\frac{R}{6}.
\end{equation}

The calculation for $[a_2]$ is slightly more tedious. It is given by 
\begin{equation} \label{simplea2.1}
\begin{aligned}
\left[a_2\right] & =\sum_l(1+l)^{-1}\bra{0}\mathcal{P}\ket{l}\bra{l}\mathcal{P}\ket{0} \\
&=\bra{0}\mathcal{P}\ket{0}^2+\frac{1}{2}\bra{0}\mathcal{P}\ket{1}\bra{1}\mathcal{P}\ket{0}+\frac{1}{3}\bra{0}\mathcal{P}\ket{2}\bra{2}\mathcal{P}\ket{0}\\
&=[Z]^2+\frac{1}{2}\left[y^m\right]\left[\partial_mZ\right]+\frac{1}{3}\left[X^{mn}\right]\left[\partial_m\partial_nZ\right] \\
&=\left(\mathcal{Q}+\frac{R}{6}\right)^2+[\phi^m][\partial_mZ]+\frac{1}{3}[g^{mn}][\partial_m\partial_nZ].
\end{aligned}
\end{equation}
The second term vanishes as $[\phi_m]=0$, for the last term, 
\begin{equation} \label{simplea2.2}
\begin{aligned}
\left[g^{mn}\right][\partial_m\partial_nZ]=[\partial_n\partial^nZ]
=&\Box{\mathcal{Q}}+[\partial_n\partial^n\phi_m\phi^m]+[\partial_n\partial^n\partial^m\phi_m]\\
&+\frac{1}{2}[\partial_n\partial^n\partial_m\partial^mB]-\frac{1}{4}[\partial_n\partial^n(\partial_mB\partial^mB)].
\end{aligned}
\end{equation}
From \eqref{phinormalexp}, we know 
\begin{equation} \label{phibrakcet2}
[\partial_n\partial^n\phi_m\phi^m]=2[\partial_n\phi_m][\partial^n\phi^m]=\frac{1}{2}F_{nm}F^{nm}, \quad [\partial_n\partial^n\partial_m\phi^m]=0.
\end{equation}
The second equation is from the fact the gauge condition is equivalent to any symmetrized partial derivatives of $\phi_m$ vanishes:
\begin{equation} \label{phiFSgauge2}
\partial_{(n_1}\dots\partial_{n_k}\phi_{m)}=0.
\end{equation}
The normal coordinate expansion of the vielbein gives the following result \cite{vandeVen97}:
\begin{equation} \label{Bbraket2}
[\partial_n\partial^n\partial_m\partial^mB]=\frac{2}{5}\Box R+\frac{2}{45}R_{mn}R^{mn}+\frac{1}{15}R_{mnpq}R^{mnpq},
\end{equation}
\begin{equation} \label{Bbraket3} [\partial_n\partial^n(\partial_mB\partial^mB)]=2[\partial_n\partial_mB][\partial^n\partial^mB]=\frac{2}{9}R_{mn}R^{mn}.
\end{equation}
Combining everything, we have the final result
\begin{equation} \label{simplea2.3}
\begin{aligned}
\left[a_2\right]=&\left(\mathcal{Q}+\frac{R}{6}\right)^2+\frac{1}{3}\Box\mathcal{Q}+\frac{1}{6}F_{mn}F^{mn}+\frac{1}{15}\Box R \\ &-\frac{1}{30}R_{mn}R^{mn}+\frac{1}{30}R_{mnpq}R^{mnpq}.
\end{aligned}
\end{equation}

The above result agrees with De Witt's original calculation. In principle, it is possible to compute higher order coefficients using the same method, as Avramidi calculated $[a_3]$ and $[a_4]$ with this machinery.

\end{document}